\documentclass[%
 aip,
 amsmath,amssymb,
 reprint,%
]{revtex4-2}

\usepackage[bookmarks=true,
            bookmarksopen=true,
            bookmarksnumbered=true,
            colorlinks=true,
            linkcolor=blue,
            citecolor=blue,
            urlcolor=blue]{hyperref}

\usepackage{graphicx}
\usepackage{dcolumn}
\usepackage{bm}
\usepackage{subcaption}

\usepackage[utf8]{inputenc}
\usepackage[T1]{fontenc}
\usepackage{newtxtext}
\usepackage{newtxmath}
\usepackage{etoolbox}
\usepackage{enumitem}
\usepackage{microtype}
\usepackage{siunitx}

\makeatletter
\def\@citess#1{\leavevmode\unskip\penalty\@M\ [\@cite{#1}{}]}
\makeatother

\makeatletter
\def\@email#1#2{%
 \endgroup
 \patchcmd{\titleblock@produce}
  {\frontmatter@RRAPformat}
  {\frontmatter@RRAPformat{\produce@RRAP{*#1\href{mailto:#2}{#2}}}\frontmatter@RRAPformat}
  {}{}
}%
\makeatother
\begin{document}

\preprint{AIP/123-QED}

\title[Effects of transitional orbit magnetization on transport and current in Z pinches]{Effects of transitional orbit magnetization on transport and current in Z pinches}
\author{Daniel W. Crews}
\email{daniel.crews@zap.energy}
\affiliation{Zap Energy Inc., Everett, WA 98203, USA}
\author{Eric T. Meier}
\affiliation{Zap Energy Inc., Everett, WA 98203, USA}
\author{Uri Shumlak}
\affiliation{Zap Energy Inc., Everett, WA 98203, USA}
\affiliation{University of Washington, Seattle, WA 98195, USA}

\date{\today}

\begin{abstract}
  The azimuthal self-magnetic field of the ideal Z pinch contains a central magnetic null.
  Trajectories around this null govern transport in the core.
  Particles follow cyclotron orbits when the guiding-center approximation holds.
  Approaching the field null, where the ordinary guiding-center regime breaks down, particles exhibit trajectories
  called, in some historical contexts, betatron orbits.
  We quantify transitional magnetization between cyclotron and betatron orbits by a magnetization parameter 
  that decomposes phase space into these orbit regimes.
  Considering the distribution of all orbits, this phase-space decomposition reveals a transitional magnetization region wherein both populations coexist.
  Classical magnetized transport theory fails within this region,
  where the diamagnetic drift reverses.
  The drift flux is instead supported by the flux of betatron orbits.
  Kinematic diffusivity remains approximately constant rather than diverging at the null.
  These transport modifications are governed solely by the number density per unit length in the ideal pinch.
\end{abstract}

\maketitle

\section{\label{sec:intro}Introduction}
Transport phenomena dictate equilibrium profiles, energy confinement,
and plasma lifetime in magnetically confined plasmas.
The magnetic null of the Z-pinch plasma complicates classical magnetized transport theory
due to incomplete magnetization of orbits.
Recent flow Z-pinch experiments aimed at fusion-relevant conditions~\cite{shumlak_fusion2020},
namely the Fusion Z-Pinch Experiment (FuZE)~\cite{zhang_2019, levitt_temperatures, Ryan_2025},
have demonstrated operation in a transitional magnetization regime.
While electrons are magnetized for the great majority of the volume,
data suggests, based on Larmor radius estimates, as much as $30\%$ of ions follow unmagnetized orbits~\cite{goyon_pressures}.
Incomplete ion magnetization challenges traditional transport models that
assume small ion gyroradii~\cite{braginskii1965}.

Several phenomena beyond traditional transport theory emerge in the transitional regime where guiding-center theory does not apply:
the diamagnetic flux is reversed
relative to the overall drift flux; the drift flux is carried by a population whose characteristic frequency is $(v_z/c)\omega_p$
and does not radially drift at the $\vec{E}\times\vec{B}$ velocity;
and the cross-field diffusivity remains nearly constant rather than diverging at the magnetic null.

This work explores these phenomena by clarifying the transitional magnetization of orbits in an azimuthal magnetic field
and providing bounds in phase space for magnetization. 
We refer to the magnetized, guiding-center orbits as cyclotron orbits.
The remaining, non-cyclotron orbits are known in the literature by various names, such as non-adiabatic orbits~\cite{haines_1978, Rostoker01121996}.
This work classifies all azimuthal field, non-cyclotron orbits as betatron orbits, based on their shared 
limiting characteristic frequency~\cite{weinberg_beam, rostoker_1994, guo_2015}, which we derive in Section~\ref{subsubsec:zero_larmor}.
These categorizations are merely idealized limits of a transitional magnetization spectrum, but the dichotomy
is a useful construction as it enables simplified theories.

Two criteria for a cyclotron orbit are derived:
(i) a bound on canonical momentum and (ii) a bound on energy.
These bounds are applied to the canonical distribution function, the Bennett solution~\cite{bennett_1934}, to
uncover a transitional magnetization region enveloping the magnetic null.
The classical theory of magnetized plasma, covering aspects of Z-pinch physics such as kinetic instabilities,
plasma-material interactions, axial shear flow and viscosity, and the equilibrium profile development of current density,
is modified in this transitional region.

This study elucidates how current is conducted in a high-beta plasma,~\textit{i.e.},{} the nature of its diamagnetic flows, 
which constitutes a large part of the particle and energy fluxes, including those leaving the system through end losses.
It is found that the cyclotron and betatron orbit subpopulations stream superthermally,
which suggests a kinetic instability drive between the magnetized and unmagnetized constituents. 
Budker's parameter is recovered as the dimensionless number governing transitional magnetization,
which for given particle mass and charge depends on the linear plasma density.
While linear density is already understood as a control parameter when 
flow Z pinches are modeled as compressible MHD flows~\cite{Datta_2024, crews_kadomtsev},
this work deepens its significance to govern pinch core transport physics.
Primarily, this work clarifies how the small Larmor radius approximation breaks down in the core of a Z pinch.

The paper structure is as follows:
Section~\ref{sec:decomposition} considers transitional magnetization, beginning in Section~\ref{subsec:hybrid_betatron}
with an examination of transitional magnetization of general particle trajectories in an azimuthal magnetic field,
while Section~\ref{subsec:partition_conditions} expresses the magnetization conditions in constants-of-motion space.
Section~\ref{sec:diamagnetic_region} applies the magnetization conditions to the distribution function,
exploring throughout Sections~\ref{subsec:density_control}-\ref{subsec:diffusion} the core-enveloping transitional magnetization region,
its scaling with a dimensionless value known as the Budker parameter, and considerations towards drift and diffusion.
Section~\ref{sec:conclusion} concludes with a discussion.

\section{\label{sec:decomposition}General motion in an azimuthal magnetic field}
Within an azimuthal magnetic field, the canonical picture of charged particle motion involves cyclotron
motion in the periphery and free-streaming motions along the axis.
Between these limits lies a transition where particles are still radially confined but have a ``gyroradius'' large enough for them to cross the axis.
Such axis-crossing orbits are known as betatron orbits in the field-reversed configuration (FRC) community~\cite{rostoker_1994}
in analogy to the terminology for beam oscillations in a quadrupole focusing field~\cite{weinberg_beam}.

To clarify this transition, we begin by considering the general motion of a charged particle in an azimuthal magnetic field.
Recall that guiding-center theory is found by an expansion in small Larmor radius $\rho$, \textit{i.e.}{},
in powers of $\rho/L_\nabla$ where $L_\nabla$ is the gradient scale length~\cite{northrop_1961, Jacobson_2025}.
We start with zero Larmor radius orbits and progressively go to first and second-order in the ratio $\rho/r$
with $r$ the distance to the pinch center,
thereby recovering the guiding-center drifts with the details contained in Appendix~\ref{app:guiding_center}.
Large Larmor radius orbits $\rho>r$ are then considered in a special analytic case detailed in Appendix~\ref{app:analytic}.

In the transitional regime, general particle motion is characterized by the two basic frequencies:
specifically, the betatron frequency $\omega_\beta$ and the cyclotron frequency $\omega_c$.
The frequency ratio $\omega_c/\omega_\beta$ is related to the constants of motion in a way
that partitions phase space into regions of qualitatively different motion,
providing clear criteria for the guiding-center gyromotion required by
traditional transport theory.

Orbits exhibiting the betatron frequency are found in plasmas with field nulls such as the
field-reversed configuration (FRC)~\cite{rostoker_1994, rostoker_2002, steinhauer_2020} and the Z pinch~\cite{haines_1978},
and arise in the theory of charged particle beams~\cite{weinberg_beam}.
Betatron orbits are a cylindrical analog of the Speiser orbits 
in the neutral line of a current sheet~\cite{speiser_1965a, sonnerup_1971, george_2020};
the meridional $(r,z)$ subclass of orbits in the cylinder are, close to the axis, identical to the Speiser
orbits of a Cartesian current sheet with zero guide field.
Betatron orbits are also known as non-adiabatic orbits for historical reasons, as none of their adiabatic invariants
coincide with the magnetic moment.

\subsection{\label{subsec:hybrid_betatron}Transition from unmagnetized to magnetized orbits}
We explore orbit magnetization with progressive generality, beginning in Section~\ref{subsubsec:zero_larmor}
with zero Larmor radius. Section~\ref{subsubsec:intro_linearized} considers the small Larmor radius orbit
and the cyclotron and betatron orbits as limiting cases. 
Section~\ref{subsubsec:finite_larmor_radius} examines finite Larmor radius corrections and compares to guiding-center theory.
Section~\ref{subsubsec:large_larmor_radius} discusses large Larmor radius in a special case.
Section~\ref{subsubsec:magnetization_parameter} introduces the magnetization parameter.

The Z pinch is an axisymmetric configuration in cylindrical coordinates $(r,\theta,z)$,
consisting of an axial current $\vec{\jmath}=j_z(r)\hat{z}$
that generates an azimuthal magnetic field $\vec{B}=B_\theta(r)\hat{\theta}$ through the vector
potential $\vec{A}=A_z(r)\hat{z}$, as expressed by $\nabla^2 A_z = -\mu_0j_z$ and $B_\theta=-\frac{dA_z}{dr}$.
Quasineutrality and sub-relativistic drift velocities ($v_z\ll c$) are assumed.
The electric field is assumed to be reducible, meaning the lab frame may be transformed to the
zero-radial-electric-field frame by a magnetoquasistatic Lorentz transformation.
That is, with primes indicating a new frame, $\vec{E}'=-\vec{v}'\times\vec{B}=0$~\cite{yoon_2025}.

The magnetoquasistatic approximation is appropriate for sub-relativistic thermal, flow, and drift speeds, 
and the reducible electric field ($\vec{E}'=-\vec{v}'\times\vec{B}=0$)
holds for shear-flow electric fields weak compared to the bulk motional field in the lab frame.
Further assumptions behind the reducible electric field are: the relative electron-ion drift is
radially uniform, the plasma species are isothermal, there are only two plasma species,
the ions are singly charged, and resistivity is neglected.

The reducible electric field approximation is employed to focus on the essential transition from weak-to-strong self-fields,
although irreducible electric fields do characterize important cases such as the skin-current pinch.
This approximation essentially limits the analysis to the canonical distribution (the Bennett solution).
However, the results obtained depend perspicuously on only a single parameter.

The small Larmor radius orbits are analyzed
using the effective potential method (detailed in Appendix~\ref{app:method}).
In this method, the radial motion is described by a one-dimensional Hamiltonian $H=K_r + V$
with $K_r$ radial kinetic energy
and $V$ the effective potential
\begin{equation}\label{eq:effective_potential}
  V = \frac{(P_z-qA_z)^2}{2m} + \frac{1}{2m}\Big(\frac{L_\theta}{r}\Big)^2
\end{equation}
where $P_z = mv_z + qA_z$ is the axial canonical momentum and $L_\theta = mrv_\theta=mr^2\omega_\theta$
is the angular momentum where $\omega_\theta$ is the azimuthal angular frequency.
The potential consists of magnetic and centrifugal terms.

\subsubsection{\label{subsubsec:zero_larmor}Zero Larmor radius orbits and the betatron frequency}
This section observes that zero Larmor radius orbits circulate the axis at the betatron frequency.
The orbit is found to be a curvature drift.
The betatron frequency is shown to be a collective property of the current-carrying plasma,
and we note how this frequency characterizes electromagnetic modes.

\begin{figure}[thbp]
  \centering
  \includegraphics[width=0.71\linewidth]{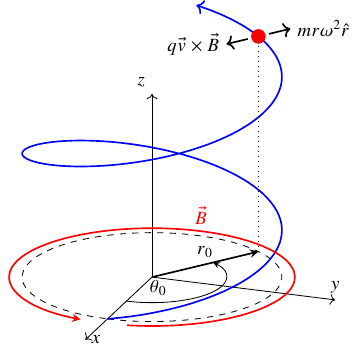}
  \caption{
    Around an O-type magnetic null, the balance of magnetic and centrifugal forces yields a helical
    orbit with a circulation rate of the betatron frequency, $\omega_\beta=\sqrt{\omega_cv_z/r}$.}
  \label{fig:helicalmotion}
\end{figure}

Magnetic and centrifugal forces balance ($\frac{dV}{dr}=0$) when $mr\omega_\theta^2 = qv_zB_\theta$,
for which the trajectory encircles the $\hat{z}$-axis at rate $\omega_\theta$ and drifts with velocity $v_z$.
Figure~\ref{fig:helicalmotion} depicts this motion.
This circulation rate $\omega_\theta=\omega_\beta \equiv \sqrt{\omega_c v_z/r}$ is
known as the betatron frequency~\cite{weinberg_beam},
where $\omega_c=qB_\theta/m$ is the cyclotron frequency.
The betatron frequency arises from the coupling of magnetic
and inertial accelerations. 
Both left and right-hand polarized circulations ($\omega_\theta=\pm\omega_\beta$) are solutions.

The zero-Larmor-radius orbit is essentially a pure curvature drift,
as the betatron frequency $\omega_\beta = \sqrt{\omega_c v_z/r}$ is equivalent to
the curvature drift expression
$\vec{v}_c = \frac{2K_\theta}{qB}\frac{\vec{R}_c\times\vec{B}}{R_c^2B}$
where $\vec{R}_c=r\hat{r}$ is the radius of curvature vector, having used $K_\theta = \frac{1}{2}mr^2\omega_\theta^2$.

The betatron frequency is a collective property of a current-carrying plasma. 
Specifically, the betatron frequency emerges from the
collective drift motions generating axial current.
In the zero-temperature limit, all particles undergo this curvature drift,
each circulating the axis with zero-mean, random polarization~\cite{weinberg_beam}.
Moving together with a radially uniform drift velocity,
these orbits carry a total current $I(r)=qnv_z\pi r^2$ where $n$ is the number density.
Under these conditions, the individual betatron frequency $\omega_\beta$ is  
tied to the collective plasma frequency $\omega_p$ by the relativistic factor $v_z/c$,
\begin{equation}\label{eq:betatron_frequency}
  \omega_\beta = \frac{v_z}{c}\omega_p.
\end{equation}
This relationship illustrates how the betatron frequency 
is intimately connected with the collective plasma properties. 

Indeed, the betatron frequency describes the characteristic frequency of collective electromagnetic phenomena.
To quote R.~C.~Davidson~\cite{davidson1990introduction}, ``the term ... proportional to $(v/c)^2\omega_p^2$ arises from [an] electromagnetic correction.''
It is well known that the electromagnetic filamentation instability, also known as the Weibel instability, grows at this frequency~\cite{davidson_1972, krall1973principles, Crews_Shumlak_2024}.
Further, the betatron frequency is thermally paired to the characteristic length of the Vlasov-Ampere system, \textit{i.e.},{} $\delta = (c/v_z)\lambda_D$ where
$\lambda_D$ is the Debye length, in the sense that $\omega_\beta\delta = \omega_p\lambda_D = v_t$ with $v_t$ the thermal velocity.
It was pointed out by S.~M.~Mahajan that with $v_z$ the relative drift velocity,
$\delta$ is the characteristic radius of a self-pinched plasma~\cite{mahajan1989exact}.

\subsubsection{\label{subsubsec:intro_linearized}Transitional magnetization of small Larmor radius orbits}
This section studies small Larmor radius orbits, 
employing linear analysis to reveal a hybrid motion involving both the cyclotron and betatron frequencies.
Transitional magnetization is found to depend on the ratio of 
these two characteristic frequencies.
This frequency ratio determines whether the orbit follows standard guiding-center theory or deviates from it.
Standard guiding-center motion emerges as a special case of a more general behavior.

Let the equilibrium quantities be denoted as $r_0$, $v_{z0}$, etc.
The frequency of infinitesimal oscillations about this equilibrium is determined by $\omega^2 = m^{-1}\frac{d^2V}{dr^2}|_{r=r_0}$.
These oscillations correspond to a finite radial temperature.
A short calculation reveals three frequency components, with $x^\prime \equiv dx/dr|_{r=r_0}$,
\begin{equation}\label{eq:oscillation_frequency}
  \omega^2 = \omega_c^2 + v_{z0}\omega_c' + 3\omega_\beta^2,
\end{equation}
in which the terms respectively indicate Larmor gyration in the local magnetic field, a field gradient-drift coupling,
and a centrifugal effect from angular momentum conservation.
The contributions $\omega_c^2$ and $v_{z0}\omega_c'$ arise from the effective magnetic
potential while $3\omega_\beta^2$ stems from the centrifugal potential.

The betatron frequency arises not only from the centrifugal interaction
but also from the gradient-drift coupling because $v_{z0}\omega_c' = \omega_\beta^2 \frac{d\ln B_\theta}{d\ln r}$.
Due to this coupling, the betatron frequency also characterizes null-crossing meridional orbits,
\textit{i.e.},{} non-encircling trajectories confined to the $(r,z)$ plane.
The gradient-drift coupling transitions from $\frac{d\ln B_\theta}{d\ln r}=+1$ in the core,
to $\frac{d\ln B_\theta}{d\ln r} = 0$ at the maximum field, and to $\frac{d\ln B_\theta}{d\ln r} = -1$
in the periphery where the field approaches its vacuum drop-off.
Because of this drift-gradient coupling, the betatron frequency characterizes the
Speiser orbits of the planar current sheet, or the transverse bounce orbits of the Weibel instability.

The centrifugal and gradient-drift terms can be combined to express a general hybrid frequency,
\begin{equation}\label{eq:exact_hybrid}
  \omega^2 = \omega_c^2 + 4F\omega_\beta^2
\end{equation}
where $F \equiv (3 + \frac{d\ln B_\theta}{d\ln r})/4$ captures the spatial variation of the field.
Equivalently, $F$ measures the concentration of electric current as, using Amp\`{e}re's law,
$F = \big(1 + j_z/\langle j_z\rangle\big)/2$
where $j_z$ is local current density and $\langle j_z\rangle = S^{-1}\int\vec{\jmath}\cdot d\vec{S}$
is current density averaged up to $r$.

The current distribution factor $F$ is order unity for typical current profiles.
For example, the factor $F\in [1, 1/2]$ (unity around the axis and one-half at large radius)
for an everywhere positive center-peaked current profile bounded between a finite central
current $j_z(0)=j_0$ and an edge $j_z\to 0$ at $r>0$.

\begin{figure}[b]
  \centering
  \includegraphics[width=\linewidth]{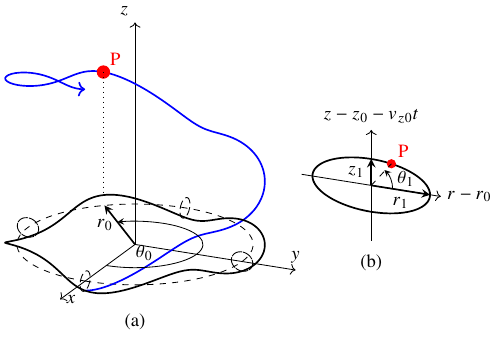}
  \caption{Illustration of the coupled cyclotron-betatron orbit in the small Larmor radius approximation:
    (a) drifting trajectory in blue
    and drift-frame trajectory in solid black, with deferent circle and bounding ellipse in dashed lines,
    (b) elliptical motion about the equilibrium radius in the drift frame (the dashed ellipse in (a)).
    A rational case $\omega=4\omega_\beta$ is shown in which the
    projected motion closes on itself.}
  \label{fig:perturbed_motion2}
\end{figure}

\begin{figure*}[htpb]
  \centering
  
  \def\rZero{1.25}       
  \def\rOne{0.075}        
  \def\omegaZero{1.0}   
  
  \begin{subfigure}{0.3\textwidth}
    \centering

      

    \includegraphics[width=\linewidth]{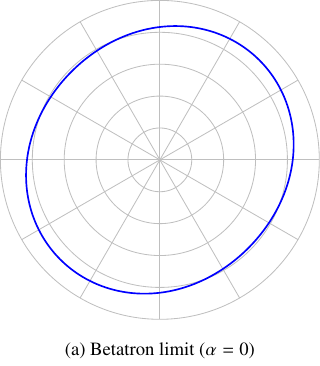}
  \end{subfigure}
  \hfill
  \begin{subfigure}{0.3\textwidth}
    \centering
      


    \includegraphics[width=\linewidth]{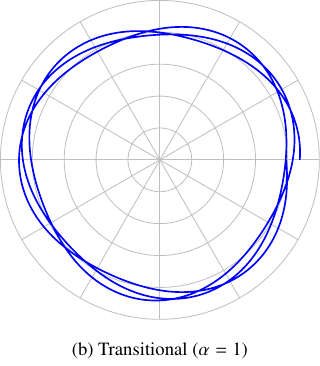}
  \end{subfigure}
  \hfill
  \begin{subfigure}{0.3\textwidth}
    \centering

      

    \includegraphics[width=\linewidth]{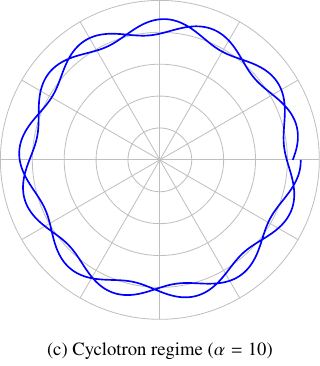}
  \end{subfigure}

  \caption{Orbits projected into the $(r,\theta)$ plane for various magnetizations 
    $\alpha=\omega_c^2/4\omega_\beta^2$, plotting the perturbative approximation
    $r(\theta)=r_0 + r_1\sin(\omega t)$ with $\theta=\omega_\beta t$
    and $\omega = 2\omega_\beta\sqrt{1+\alpha}$.
    The trajectories indicate transition from (a) an axis-centered ellipse
    (the well-known perturbed beam orbit), (b) the transitional regime, and (c) the guiding-center
    or magnetized regime. The parameter regime $\alpha\leq 1$ essentially does not display
    gyromotion, but has been described as precession of an elliptical orbit~\cite{weinberg_beam}.}
  \label{fig:polar_plots}
\end{figure*}

For illustration of the general motion, we now outline the transitional orbit, as shown in Fig.~\ref{fig:perturbed_motion2}.
The azimuthal circulation at frequency $\omega_\beta$ and the radial oscillation at frequency $\omega$
are described by two gyrophases,
\begin{subequations}\label{eq:gyrophases}
\begin{eqnarray}
  \theta_0 &=& \omega_\beta t + \varphi_0,\label{eq:theta0}\\
  \theta_1 &=& \omega t + \varphi_1\label{eq:theta1}
\end{eqnarray}
\end{subequations}
with the $\varphi$ arbitrary phases.
The trajectory to first-order in $\delta \equiv r(t) - r_0$ follows
by canonical momentum conservation ($P_z = mv_z + qA_z$ with $A_z = A_{z0} - B_{\theta 0}\delta + \cdots$) as
\begin{subequations}\label{eq:linear_motion}
\begin{eqnarray}
  r(t) &=& r_0 + r_1\sin(\theta_1),\label{eq:r0}\\
  z(t) &=& z_0 + v_{z0}t - z_1\cos(\theta_1)\label{eq:z1}
\end{eqnarray}
\end{subequations}
where the amplitudes are $r_0 = \big(\frac{\omega_c}{\omega_\beta}\big)^2\frac{v_{z0}}{\omega_c}$,
$r_1 = \frac{v_{r1}}{\omega}$,
$z_1=\big(\frac{\omega_c}{\omega}\big)r_1$ and $z_0$ is an axial shift.
In the drifting frame ($z' = z - z_0 - v_{z0}t$), the particle gyrates in an ellipse about the
guiding center, as shown in Fig.~\ref{fig:perturbed_motion2} for the
rational special case $\omega/\omega_\beta = 4$.
The two-phase gyration of the particle displays two mutually dependent motions:
the axis-enveloping rotation of the guiding center and the internal gyration about the guiding center;
the two are connected by the frequency ratio $\omega_c/\omega_\beta$.
Figure~\ref{fig:polar_plots} illustrates how this frequency ratio controls the characteristic motion
via the projection to the $(r, \theta$) plane.

The origin of the hybrid motion is particularly clear in the drift frame $\vec{v}' = \vec{v} - v_{z0}\hat{z}$
(assuming the magnetoquasistatic limit $v_{z0}\ll c$), in which a radially inward electric field
balances the centrifugal force.
In this frame, the equation of motion is
\begin{subequations}\label{eq:equation_of_motion}
\begin{eqnarray}
  \ddot{\delta} &=& -\omega_\beta^2\delta - \omega_c\dot{z}',\label{eq:delta_ddot}\\
  \ddot{z}' &=& \omega_c\dot{\delta},\label{eq:z_ddot}
\end{eqnarray}
\end{subequations}
describing radial oscillation at frequency $\omega_\beta$ coupled
to rotation about the azimuth at frequency $\omega_c$.

\subsubsection{\label{subsubsec:finite_larmor_radius}Guiding-center drifts at finite Larmor radius}
This section determines finite Larmor radius corrections to the orbits
by examining the motion to second order in $r_1/r_0$.
Appendix~\ref{app:guiding_center} contains the mathematical details.
The standard guiding-center description is found to emerge in the limit where the 
cyclotron frequency dominates, $\omega_c\gg \omega_\beta$.
In the cases $\omega_\beta\approx \omega_c$ and $\omega_\beta\gg \omega_c$,
the guiding-center description is significantly modified by transitional magnetization effects.

At finite Larmor radius, the $\theta_1$-averaged position, or guiding center, of the particle shifts from the radius $r_0$
because the effective potential is nonlinear.
Averaging over the period $T = 2\pi/\omega$,
the displacement $\Delta\equiv (\langle r\rangle - r_0)/r_1$ is 
\begin{equation}\label{eq:displacement}
  \Delta =
  3\Big(\frac{r_1}{r_0}\Big)\Big(\frac{\omega_\beta}{\omega}\Big)^2\Big(1 - \frac{1}{12}r_0^2\frac{\omega_c''}{\omega_c}\Big)
  - \frac{3}{4}\Big(\frac{\omega_c}{\omega}\Big)^2\Big(r_1\frac{\omega_c'}{\omega_c}\Big)
\end{equation}
where the frequencies are evaluated at the original radius $r_0$.
This expression contains two terms: 
the first term is proportional to $\omega_\beta$, arising from the centrifugal effect and the gradient-drift coupling,
and the second term is proportional to $\omega_c$, from the local magnetic field and its gradient.
Examining the limit where cyclotron motion dominates, $\omega_c\gg\omega_\beta$,
Eq.~\ref{eq:displacement} reduces to the second term which
takes the form of the standard guiding-center displacement~\cite{brizard2017},
$\Delta_\text{gc}=-\frac{3}{4}\frac{r_1}{L_\nabla}$ where $L_\nabla = \omega_c/\omega_c'$.
In this way, the betatron contribution to the motion participates in a generalized guiding-center motion.

The guiding-center drift velocity is found 
by combining canonical momentum conservation with the displacement $\Delta$.
Expanding the vector potential to second order in the amplitude $r_1$ and
averaging over $T$, as in Appendix~\ref{app:guiding_center}, leads to
\begin{equation}\label{eq:drift1}
  \langle v_z\rangle = v_{z0} + \omega_cr_1\Delta + \frac{1}{2}\omega_c'r_1^2
\end{equation}
where $v_{z0}$ is the original curvature drift of the equilibrium orbit.
Substitution of Eq.~\ref{eq:displacement} for the average displacement leads to
\begin{equation}\label{eq:drift2}
  \begin{aligned}
  \langle v_z\rangle = v_{z0}\Big(1 + 6\Big(\frac{\omega_\beta}{\omega}\Big)^4&\frac{\langle K_r\rangle}{K_{z0}}
  \Big(1 - \frac{1}{12}r_0^2\frac{\omega_c''}{\omega_c}\Big)\Big)\\
    &- \frac{\langle K_r\rangle}{m}\frac{\omega_c'}{\omega^2}\Big(3\Big(\frac{\omega_c}{\omega}\Big)^2-1\Big)
  \end{aligned}
\end{equation}
where $\langle K_r\rangle$ is the gyro-averaged radial kinetic energy.
The two essential effects are a modification to the curvature drift
proportional to $\omega_\beta/\omega$, and a gradient-dependent drift proportional to $\omega_c/\omega$.
In the limit $\omega\to\omega_c$,
standard guiding-center theory emerges as $\langle v_z\rangle \to v_c + v_\nabla$ (\textit{i.e.},{} sum of curvature and grad-$|\vec{B}|$ drifts)
where $v_\nabla\equiv -\frac{K_\perp}{m}\frac{\omega_c'}{\omega_c^2}$ and $K_\perp = 2\langle K_r\rangle$.
Clearly, the guiding-center drift is modified by transitional magnetization,
which is controlled by the frequency ratio $\omega_c/\omega_\beta$. 

\subsubsection{\label{subsubsec:large_larmor_radius}Large Larmor radius orbits in a constant gradient}
In this section, we show that transitional magnetization of large Larmor radius orbits, in a special case,
remains controlled by the frequency ratio $\omega_c/\omega_\beta$.
This special case is an analytic solution in uniform current density $\vec{\jmath}=j_0\hat{z}$.
This cylindrical solution, and a generalization to elliptic field lines, was first discussed by Kim and Cary~\cite{kim_1983}.
The planar analog is often used to analyze the Speiser orbits in current sheets~\cite{speiser_1965a, sonnerup_1971, brizard2017}.

As shown in Appendix~\ref{app:analytic}, the general solution 
is given by
\begin{equation}\label{eq:radial_solution_text}
  r^2(t) = r_0^2 - r_1^2\text{sn}^2(\widetilde{\omega} t | m)
\end{equation}
where $\text{sn}$ is the Jacobi elliptic sine, $\widetilde{\omega}$ is a
characteristic frequency (distinct from the previous section) and $r_1$ a characteristic amplitude.
The elliptic modulus $m=m(\omega_c/\omega_\beta, K_\theta/K_z)$ depends on the frequency ratio
and the kinetic energies parallel and perpendicular to the magnetic field.

Computing the drift velocity,
averaging over the exact gyroperiod, and expanding to leading order in $\omega_\beta/\omega_c$ gives 
\begin{equation}\label{eq:guiding-center-spin}
  \langle v_z\rangle = v_c + v_\nabla + \mathcal{O}((\omega_\beta/\omega_c)^4)
\end{equation}
where $v_c$ is curvature drift, $v_\nabla$ is the $\nabla|\vec{B}|$ drift.
The guiding-center drift is modified by transitional magnetization effects.
This result indicates insensitivity to the small Larmor radius assumption of Section~\ref{subsubsec:finite_larmor_radius}.

Additional current profile effects arise for orbits which sample a large range of magnetic field,
such as a large amplitude orbit which experiences both $\omega_c'<0$ and $\omega_c'>0$.
The solution analyzed here does not capture such gradient effects
because the trajectory is assumed to traverse a constant gradient.
Such current profile effects can qualitatively modify the drift properties 
of very high energy orbits.

\subsubsection{\label{subsubsec:magnetization_parameter}The magnetization parameter}
The previous sections have shown that transitional magnetization is controlled by the relative betatron
and cyclotron frequencies of the orbit.
This section casts the frequency ratio as a magnetization parameter and expresses it in an approximate form suitable for
phase space decomposition.

The frequency ratio $\omega_c/\omega_\beta$ essentially measures the relative rates
of the two gyrophases $\theta_1$ and $\theta_0$. 
When the cyclotron frequency exceeds the betatron frequency, $\omega_c\gg\omega_\beta$,
the particle completes many gyroperiods before encircling the axis, and hence displays guiding-center behavior.
Meridional trajectories, confined to the $(r,z)$ plane, follow a similar logic: orbits only cross the axis 
if the Larmor gyration is outpaced by the betatron oscillation about the null.

Orbit magnetization is parametrized by
\begin{equation}\label{eq:alpha_def}
\alpha = \frac{\omega_c^2}{4\omega_\beta^2},
\end{equation}
which quantifies the dominant frequency of the unified motion given by Eq.~\ref{eq:exact_hybrid}.
There is a local, current profile-dependent, order-unity correction to the factor $4$ in Eq.~\ref{eq:alpha_def} (the $F$ of Eq.~\ref{eq:exact_hybrid}).
This parameter is essentially equivalent to the one often used in reconnection theory~\cite{buchner_1989, seiji_2016}.

We now reformulate the magnetization parameter in terms of the flux function as follows,
\begin{equation}\label{eq:inductance_form}
  \alpha = \Big(\frac{\mu_0/4\pi}{L'}\Big)\frac{q\psi^\prime_\theta}{2mv_z}
\end{equation}
where $\psi^\prime_\theta = -A_z = L'I$ is azimuthal flux per unit length, $I = \mu_0^{-1}2\pi rB_\theta$
is enclosed current, and $L^\prime = \psi^\prime_\theta/I$ is inductance per unit length (all as functions of radius). 
Typically, the inductance $L^\prime$ differs from $\mu_0/4\pi$ only logarithmically in $r$. 
Therefore,
\begin{equation}\label{eq:trapping_parameter}
  \alpha\approx -\frac{qA_z}{2mv_z}
\end{equation}
up to logarithmic corrections in the inductance and an order-unity correction in the current profile.
Physically speaking, Eq.~\ref{eq:trapping_parameter} considers only the self-flux $\psi^\prime_{\text{self}}=\frac{\mu_0}{4\pi}I$.
This approximate form of $\alpha$ relates the canonical momentum to either mechanical momentum or potential momentum,
\begin{equation}\label{eq:alpha_forms}
  P_z = (1-2\alpha)mv_z = \Big(1 - \frac{1}{2\alpha}\Big)qA_z.
\end{equation}

\subsection{\label{subsec:partition_conditions}Domain of magnetization in phase space}
This section utilizes the magnetization parameter in the form involving the vector potential, Eq.~\ref{eq:trapping_parameter},
to identify the domain of canonical phase space $(P_z, H)$ that is magnetized by the azimuthal magnetic field.
The condition $\alpha > 1$ leads to,
\begin{subequations}\label{eq:hp_limits}
\begin{eqnarray}
  \frac{(P_z-qA_z)^2}{2m} &< H <& \frac{P_z^2}{2m}, \label{eq:klim}
  \\
  -\infty &< P_z <& \frac{qA_z}{2},\label{eq:plim}
\end{eqnarray}
\end{subequations}
for a positively charged particle ($q>0$), shown as follows.
The momentum bound comes directly from Eq.~\ref{eq:trapping_parameter},
while the lower energy bound is the minimum possible particle energy
(describing zero Larmor radius orbits).
Interestingly, in the vacuum field $B_\theta\to r^{-1}$ the expression $\omega^2=\omega_c^2+2\omega_\beta^2$
suggests that the exact momentum bound limits to $P_z<0$, but around the magnetic null
where $\omega^2=\omega_c^2+4\omega_\beta^2$ the momentum bound is tied to the flux function as shown.

The upper bound on energy for magnetization is demonstrated by an analysis at the radial turning points
(where $V=H$).
First, observe that the difference between $P_z^2/2m$ and the axial kinetic energy
$K_z=\frac{(P_z-qA_z)^2}{2m}$ at the turning point is
\begin{equation}
  \frac{P_z^2}{2m} - \frac{(P_z-qA_z)^2}{2m} = 4\alpha(\alpha-1)K_z
\end{equation}
where $\alpha$ and $K_z$ are both evaluated at the turning point.
Then evaluating the effective potential at the turning point,
\begin{equation}
  V|_{r=r_t}<\frac{P_z^2}{2m} \implies \frac{K_\theta}{K_z}\Big|_{r=r_t}<4\alpha(\alpha-1).
\end{equation}
The left-hand-side must be positive, so the parameter $\alpha>1$ or $\alpha<0$.
The case $\alpha<0$ is excluded by $P_z<\frac{qA_z}{2}$.
Thus, the energy bound $H<\frac{P_z^2}{2m}$ excludes orbits with $\alpha<1$,
and so gives the energy bound for magnetization of both axis-encircling and meridional motions.
The physical meaning of the energy bound is that motions satisfying it have $\omega_c>2\omega_\beta$.

\subsubsection{Phase-space separatrix and approximation of the bounds}
For meridional ($L_\theta=0$) trajectories, the conditions $P_z=qA_z/2$ and $H_p\equiv P_z^2/2m$ mark
precisely the separatrix in phase space for any vector potential profile~\cite{yoon_2025}
(see Fig.~\ref{fig:example_orbits}).
Further, for the axis-encircling orbits, we have shown that the momentum bound $P_z<qA_z/2$
indicates where the local cyclotron contribution to the frequency exceeds the centrifugal
and drift-gradient betatron contributions, indicating applicability of the guiding-center model.
However, the current profile effects (the factor $F$ in Eq.~\ref{eq:exact_hybrid}
and the inductance per unit length in Eq.~\ref{eq:inductance_form}) ultimately make
the bounds in Eqs.~\ref{eq:hp_limits} merely a convention,
albeit a physically motivated one. 
In complete generality, such effects may substantially modify these bounds.

\begin{figure*}[htpb]
  \includegraphics[width=0.875\textwidth]{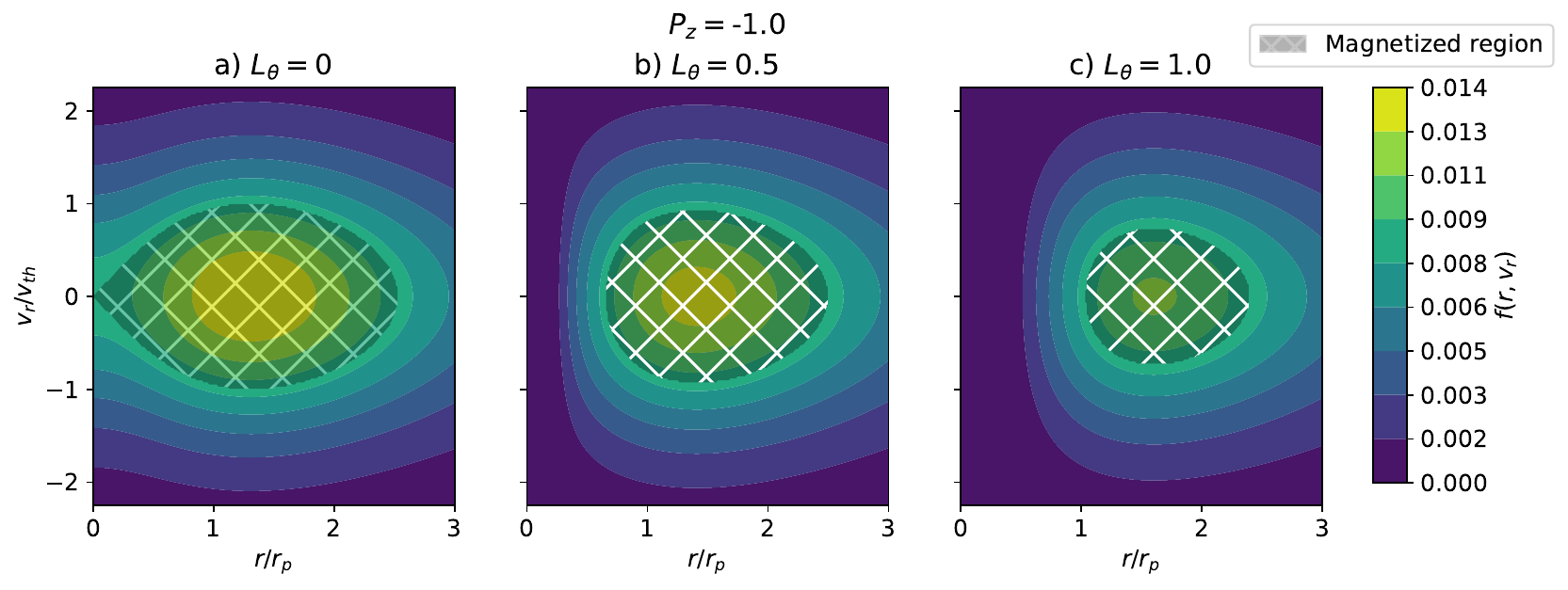}
  \includegraphics[width=0.875\textwidth]{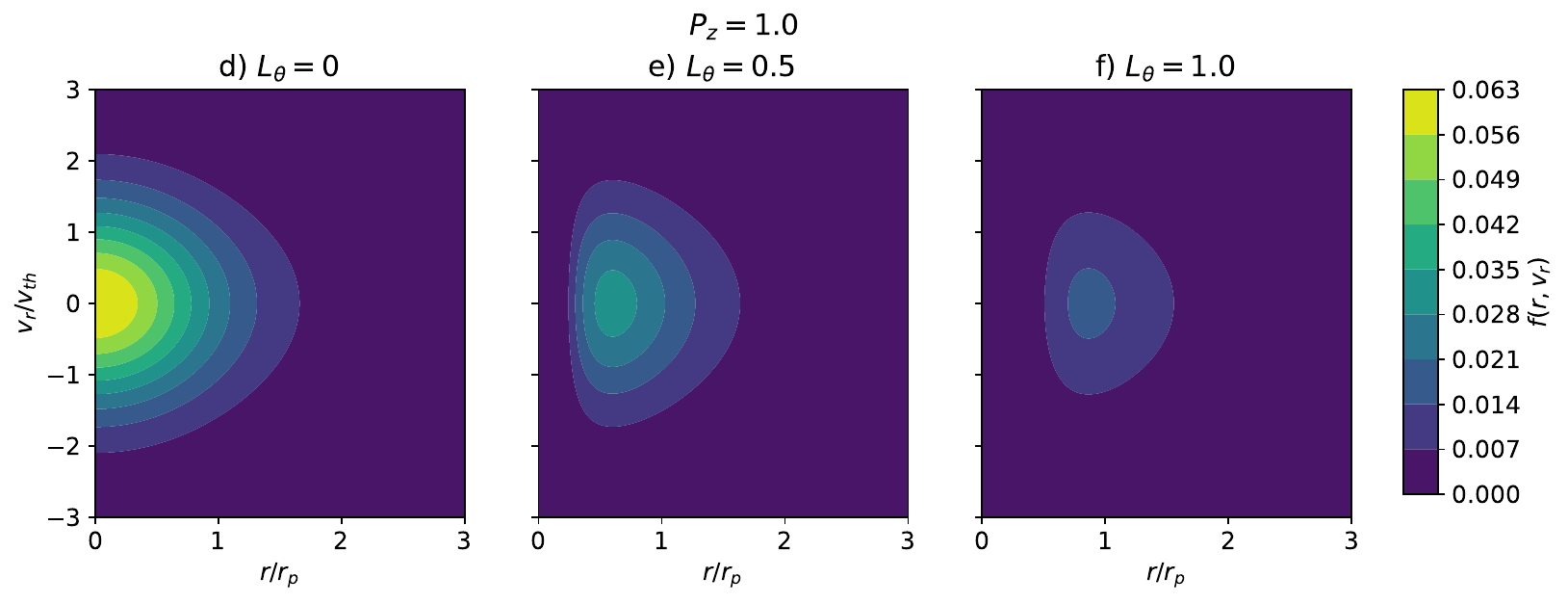}
  \caption{Phase space density $f(r, v_r)$ of Eq.~\ref{eq:canonical_distribution}
    for fixed values of canonical momentum $P_z<0$ (top row, a-c) and $P_z>0$ (bottom row, d-f),
    and angular momentum $L_\theta=0$, $0.5$, and $1.0$ (columns). The shaded region hatched in white marks the
    domain satisfying the magnetization bounds (Eqs.~\ref{eq:hp_limits}) which separates magnetized orbits
    ($H\leq P_z^2/2m$) from unmagnetized orbits.
    Only negative $P_z$ admits magnetized orbits, while for positive $P_z$ the limits
    cannot be satisfied, so no orbits are magnetized.}
  \label{fig:magnetized_phase_space}
\end{figure*}

\begin{figure*}[htpb]
  \includegraphics[width=0.875\linewidth]{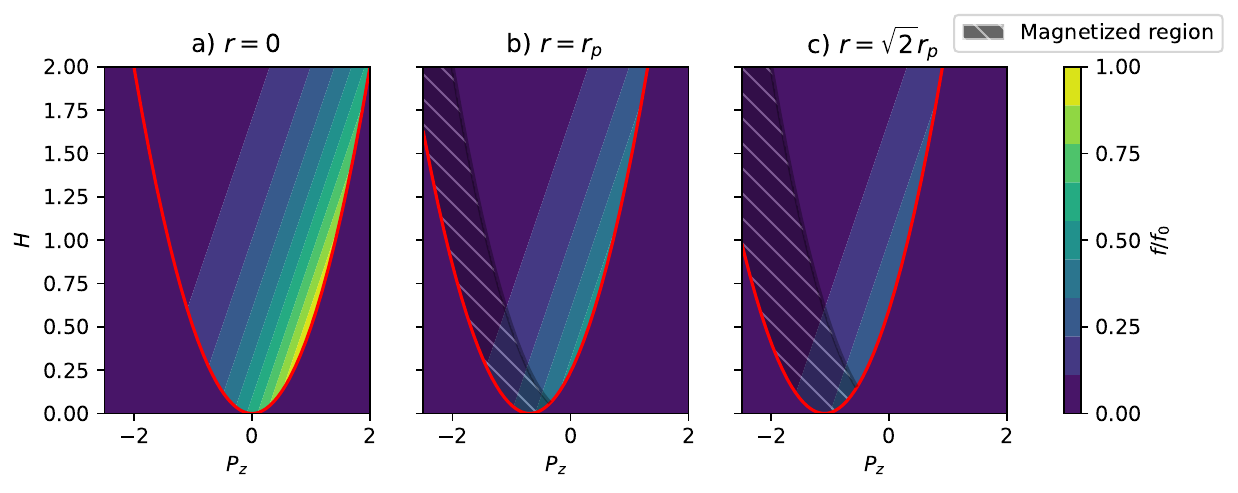}
  \caption{
    Phase space density $f(P_z,H)$ of Eq.~\ref{eq:canonical_distribution}
    at various radii: a) $r=0$, b) $r=r_p$ and c) $r=\sqrt{2}r_p$,
    normalized to the distribution's maximum at $r=0$.
    The red parabola depicts the minimum energy $H_\text{min}=(P_z-q A_z)^2/2m$
    a particle may have, thereby defining the domain $H>H_\text{min}$.
    The hatched, shaded region indicates the magnetized domain (Eqs.~\ref{eq:hp_limits}).
    Unmagnetized particles exist above the red line and outside of the magnetized region.
    All trajectories through the magnetic null ($r=0$) are meridional betatron orbits.
    As $r$ increases, progressively more of the population which thread that radius follow cyclotron orbits.
    \label{fig:energy_momentum_figure}}
\end{figure*}

\subsubsection{Transitional magnetization of the canonical distribution}
We illustrate the magnetized region of phase space for the
canonical distribution function of a Z pinch~\cite{mahajan1989equilibrium},
\begin{equation}\label{eq:canonical_distribution}
f(P_z, H) = Z^{-1}\exp(\beta u_0 P_z)\exp(-\beta H).
\end{equation}
Here $\beta=(k_BT)^{-1}$ is inverse temperature, $u_0$ is the macroscopic species axial drift,
and $Z=(\beta/2\pi)^{-3/2}e^{\beta m u_0^2/2}$ is the partition function.
Figures~\ref{fig:magnetized_phase_space} and~\ref{fig:energy_momentum_figure} illustrate how the
distribution partitions into the magnetized and unmagnetized domains in different coordinate
representations, in the specific case of the Bennett kinetic equilibrium~\cite{mahajan1989equilibrium}
for which the density and flux functions are given by
\begin{align}
  n(r) &= n_0(1 + (r/r_p)^2)^{-2},\label{eq:density}\\
  A_z(r) &= \frac{\mu_0}{8\pi}I_\infty\ln\Big(\frac{n(r)}{n_0}\Big)\label{eq:flux}
\end{align}
with $I_\infty$ the total current enclosed at $r\to\infty$, $r_p$ the characteristic radius, and $n_0$ the characteristic density.

Figure~\ref{fig:magnetized_phase_space} shows reduced-dimensional slices at various fixed values of
$(L_\theta, P_z)$ in the $(r, v_r)$ plane, with the phase space density
depicted in units normalized to the thermal state (\textit{i.e.},{} $L_\theta$ to $r_pmv_t$ and $P_z$ to $mv_t$)
assuming transitional magnetization ($u/v_t=1$). 
The density within the shaded region with white hatches satisfies the magnetization bounds,
while orbits outside of these bounds are unmagnetized.
Only negative $P_z$ admits magnetized orbits, while for positive $P_z$ the limits cannot be satisfied, so no orbits are magnetized
(cf. Section~\ref{subsubsec:magnetization_parameter}).
In constrast, Fig.~\ref{fig:energy_momentum_figure} depicts the entire phase space in $(P_z, H)$ coordinates
for three representative radii ($r=0$, $r=r_p$, and $r=\sqrt{2}r_p$) in the same transitional magnetization regime.
The shaded overlay again highlights the magnetized domain.

Ultimately, we emphasize that the mere condition $P_z<0$ is insufficient to characterize a population as
magnetized, in the sense of applicability of the guiding-center model,
lest even $P_z<0$ orbits passing through the magnetic null and with an orbit-averaged position of
$\langle r\rangle=0$ be considered cyclotron orbits.
For this reason, the energy bound $H\leq H_p$ is essential to distinguish the magnetized
from the unmagnetized population.

\section{\label{sec:diamagnetic_region}The transitional magnetization region}
The magnetization conditions of Section~\ref{subsec:partition_conditions} allow explicit calculation,
via moments of the magnetized phase space, of the boundary layer between the central magnetic null
and the magnetized periphery.
We call this boundary layer the transitional magnetization region.
This calculation clearly reveals transport properties of the region,
and is done as follows. Section~\ref{subsec:density_control} demonstrates that the canonical distribution is parametrized solely by
linear density and recalls Budker's parameter.
Section~\ref{subsec:moments} computes the partial densities and fluxes of the cyclotron and betatron orbit populations.
``Partial'' is used in the sense of partial pressures.
Section~\ref{subsec:diamagnetic} examines the drift flux, or axial transport, of the sub-populations
and looks at these fluxes in the lab frame.
Section~\ref{subsec:diffusion} concludes by estimating classical diffusivity within the region.

\subsection{\label{subsec:density_control}Parametrization of equilibrium by linear density}
The canonical distribution function and flux function are parametrized only by particle
properties (mass and charge) and the linear plasma density $N$ (particles per unit length).
It has long been recognized that $N$ measures the degree of self-magnetization~\cite{finkelstein_1959};
indeed, the transitional magnetization region vanishes as $N\to\infty$.

Parametrization by $N$ is inferable from normalization to the thermal state of that plasma species,
\textit{i.e.},{} $\widetilde{v}_s=v/v_{ts}$, $\widetilde{P}_z=P_z/m_sv_{ts}$, $\widetilde{H}=H/m_sv_{ts}^2$,
$\widetilde{A}_z = q_sA_z/m_sv_{ts}$, etc.{}, where $s$ denotes the electron or ion species.
With the species drift parameterized as $\chi_s \equiv u_{s}/v_{ts}$, 
the distribution function normalizes to
\begin{equation}\label{eq:normalized_distribution}
  \widetilde{f}_s(P_z, H; \chi_s) = \exp(-\chi_s^2/2)\exp(\chi_s\widetilde{P}_z)\exp(-\widetilde{H}).
\end{equation}
The normalization of the flux function (Eq.~\ref{eq:flux}) is expressible as electric potential in the
frame in which species force equilibrium is purely electrostatic ($\Phi_s = -\vec{v}_s\cdot\vec{A}$), namely
\begin{equation}\label{eq:normalized_voltage}
  \frac{q_s\Phi_s}{kT_s} = -\nu_s\chi_s^2\ln\Big(\frac{n}{n_0}\Big)
\end{equation}
where $\nu_s$ is Budker's parameter for species $s$,
\begin{equation}\label{eq:budker}
  \nu_s \equiv \frac{\mu_0q_s^2N_s}{4\pi m_s} = N_sr_s,
\end{equation}
where  $N_s$ is linear density and $r_s$ the classical charge radius.
Budker's parameter expresses the number of co-planar particles within a cylindrical section of length $r_s$,
and is known to measure the strength of the self-field carried by a plasma current~\cite{budker_parameter}.

Budker's parameter is the multiplicative inverse of the species drift squared, as shown by
applying the virial theorem, \textit{i.e.},{} Bennett relation, in the form $\frac{\mu_0}{8\pi}I_\infty^2 = 2N kT$
(assuming $T_e=T_i$ and $N_e=N_i$), from which a simple calculation finds
\begin{equation}\label{eq:drift_budker_relation}
  \nu_s\chi_s^2 = 1
\end{equation}
having used $u_{e}=-u_{i}$, the zero electric field frame property.
Equation~\ref{eq:drift_budker_relation} 
means that moments of the canonical distribution, normalized to the thermal state,
are parametrized solely by linear density.
In addition, Eq.~\ref{eq:drift_budker_relation} combined with Eq.~\ref{eq:normalized_voltage} produces
$n_s = n_0\exp(-q_s\Phi_s/kT_s)$, a property of thermal equilibrium.
An important consequence of the virial relation, Eq.~\ref{eq:drift_budker_relation}, is
that transitional magnetization is parametrized by linear density alone, as seen in the following section.

\begin{figure*}[thpb]
  \includegraphics[width=0.875\linewidth]{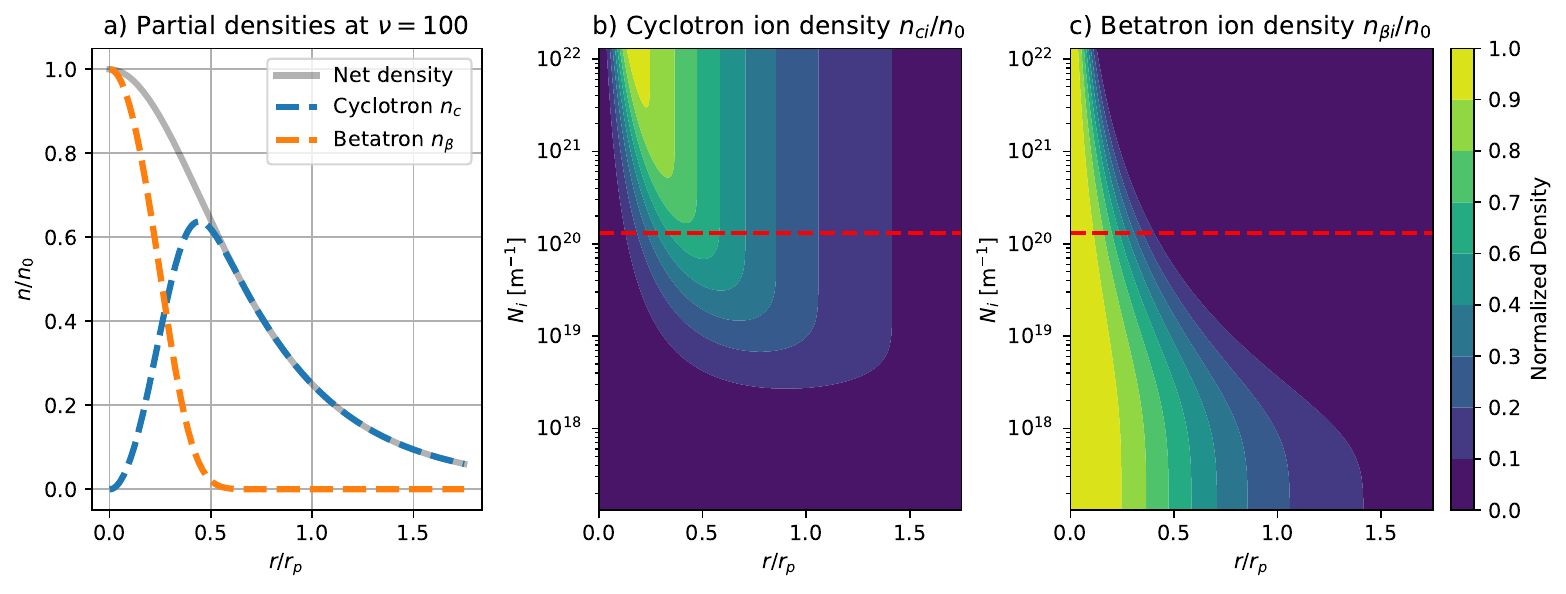}
  \caption{Densities of cyclotron ($n_c$) and betatron ($n_\beta$) trajectories 
    at a) Budker's parameter $\nu=100$, and b-c) as a function of linear density
    specialized to deuterium ions in an electron-deuterium plasma, where the dashed red line indicates $\nu_i=100$.
    At $\nu=100$, the pinch radius consists of twenty Larmor radii ($r_p=20 \rho_s$) while
    the transitional region thickness is roughly ten Larmor radii ($\Delta_s\approx 10 \rho_s$).
    The thickness in Larmor radii scales to leading order with Budker's parameter as
    $\Delta_s \approx 2\sqrt{2}\nu_s^{1/4}\rho_s$.
    \label{fig:cyclotron_densities}}
\end{figure*}

\subsection{\label{subsec:moments}Densities and fluxes of the cyclotron and betatron orbits}
This section computes the moments of the magnetized part of the canonical distribution.
The partial density of the cyclotron orbits, ``partial'' in the partial pressure sense, reveals
the relative fraction of guiding-center orbits at each radius, which delineates the magnetized regime in which reduced
models such as gyrokinetic theory or standard transport closures are applicable.
The particle flux, meanwhile, describes the proportion of the drift flux carried by the cyclotron and
betatron orbits.
These partial fluxes provide a tangible illustration of the balance between the diamagnetic
drift, emerging from the dynamics of the cyclotron orbits, and the so-called
``singular current'' of the flux of betatron orbits.

The transitional magnetization region is found to be tens of Larmor radii thick for a Z pinch,
and hence modifies ion transport significantly even at substantial linear densities.
Further, the relative drift of the subpopulations is superthermal within the transitional region,
suggesting a possible kinetic instability drive between the magnetized and unmagnetized constituents.

The moments are transformed to canonical coordinates by
\begin{eqnarray}
  \int_{-\infty}^\infty\int_{-\infty}^\infty\int_{-\infty}^\infty dv_rdv_\theta dv_z f(v_r,v_\theta,v_z)\nonumber\\
  = \int_{-\infty}^\infty dP_z \int_{H_\text{min}}^\infty dH f(P_z, H)\label{eq:zero_moment_HP}
\end{eqnarray}
where energies are restricted by $H>H_\text{min} \equiv (P_z-q_s A_z)^2/2m_s$.
The density of cyclotron orbits is found by limiting to the magnetization bounds
in Eq.~\ref{eq:zero_moment_HP} defined by Eqs.~\ref{eq:hp_limits},
\begin{equation}\label{eq:cyclo_den}
  n_\text{cs}(A_z) \equiv \int_{-\infty}^{q_s A_z/2} dP_z \int_{H_\text{min}}^{P_z^2/2m_s} dH f(P_z, H)
\end{equation}
as a flux function.
The complement density, whose characteristic frequency is the betatron frequency,
is then defined as $n_{\beta s} \equiv n_s - n_{cs}$.
For the canonical distribution and flux function (the Bennett solution), Eq.~\ref{eq:cyclo_den} gives
\begin{equation}\label{eq:split_densities}
  n_{cs}(r;\nu_s) = \frac{n_s(r)\text{erfc}_+(r; \nu_s) - n_{0s}\text{erfc}_-(r; \nu_s)}{2}
\end{equation}
where the pair of functions
\begin{equation}\label{eq:erf_def}
  \text{erfc}_\pm(r; \nu_s) \equiv \text{erfc}\Big(\frac{\chi_s}{\sqrt{2}}\Big(1 \pm \frac{1}{2}\nu_s\ln(n)\Big)\Big)
\end{equation}
are complementary error functions evaluated at the drift-shifted vector potential
(having used $q_sA_z/m_sv_{ts} = \nu_s\chi_s\ln(n)$).

The axial flux of cyclotron orbits is calculated by the first moment of velocity,
\begin{equation}
  \Gamma_\text{cs} \equiv \int_{-\infty}^{q_sA_z/2} dP_z \frac{(P_z - q_sA_z)}{m_s} \int_{H_\text{min}}^{P_z^2/2m_s} dH f(P_z, H),
\end{equation}
and evaluates for the canonical distribution to
\begin{equation}\label{eq:split_currents}
  \Gamma_{cs}(r; \nu_s) = n_{cs}(r)u_{0s} + n_{0s}u_{0s}\nu_s\ln(n)\frac{\text{erfc}_-(r; \nu_s)}{2}
\end{equation}
where $u_{0s}$ is the bulk velocity.
The complement $\Gamma_{\beta s} \equiv \Gamma_s - \Gamma_{cs}$ is the flux of betatron orbits,
where $\Gamma_s = n_{0s}u_{0s}$ is the net axial particle flux of species $s$.
The average velocity of either sub-population is simply
$\langle v\rangle_{\alpha s} = \Gamma_{\alpha s}/n_{\alpha s}$.

\begin{figure*}[htpb]
  \includegraphics[width=0.875\linewidth]{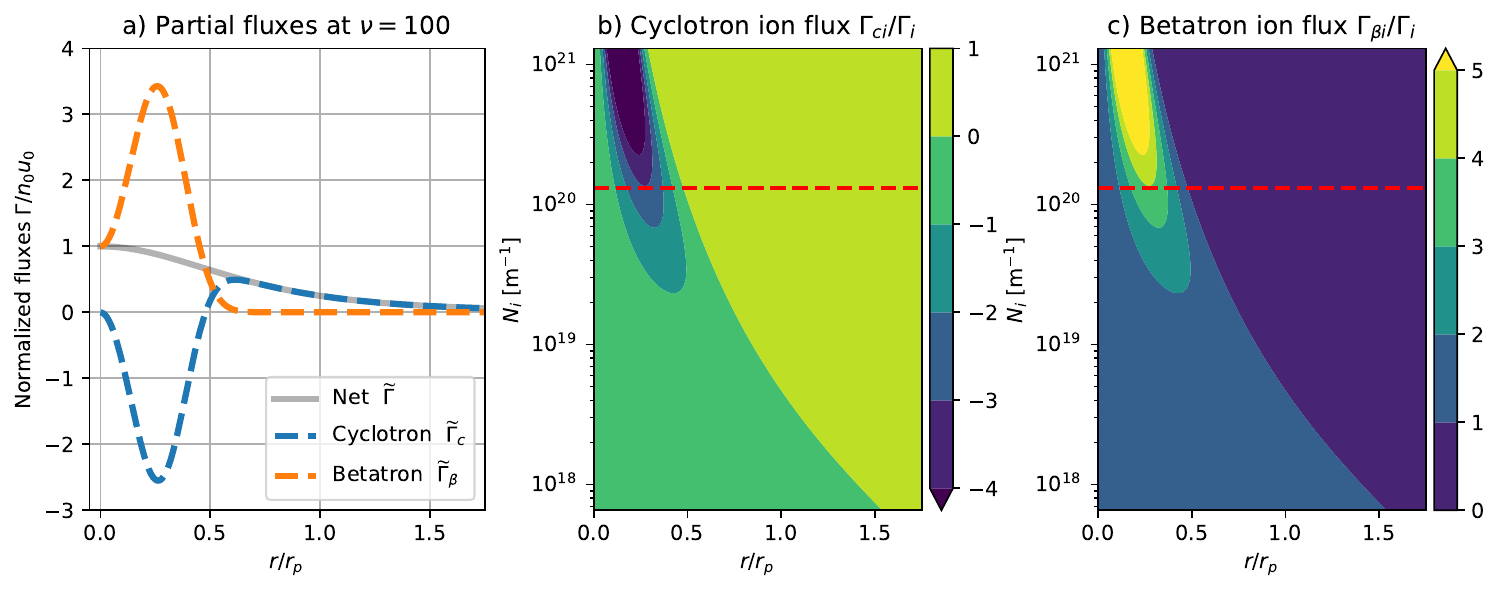}
  \caption{
    Particle fluxes $(\Gamma)$ of cyclotron ($\Gamma_c$) and betatron ($\Gamma_\beta$) orbits,
    measured in the entirely magnetic reference frame ($\vec{E}^\prime = 0$).
    Fluxes are normalized as $\Gamma/n_0u_0$ in part a) which is specialized
    to Budker's parameter $\nu=100$, while parts b-c) are normalized to 
    the local ion drift flux $\Gamma_i(r)=n_i(r)u_0$ assuming an electron-deuterium plasma.
    The dashed red line indicates $\nu_i=100$.
    Within the transitional region, a positive gradient in the density of cyclotron orbits ($n_c'>0$)
    generates a ``reversed'' diamagnetic flux (opposing the net drift flux), while the betatron orbits
    support the positive flux which carries the current.
  \label{fig:cyclotron_currents}}
\end{figure*}

Figures~\ref{fig:cyclotron_densities} and~\ref{fig:cyclotron_currents} show the density and particle flux profiles
of the cyclotron and betatron orbits, both as line-outs for Budker's parameter $\nu_s=100$, and as a function
of the ion linear density $N_i$ in the case of an electron-deuterium plasma, for which
$\nu_i \approx \SI{7.7e-19}{}N_i$ with $N_i$ in $\SI{}{m^{-1}}$.
For deuterium ions, $N_i\approx \SI{1e18}{m^{-1}}$ marks the threshold linear density for ion
magnetization for which Budker's parameter $\nu_i\approx 1$.

\begin{figure*}[htpb]
  \includegraphics[width=0.9\linewidth]{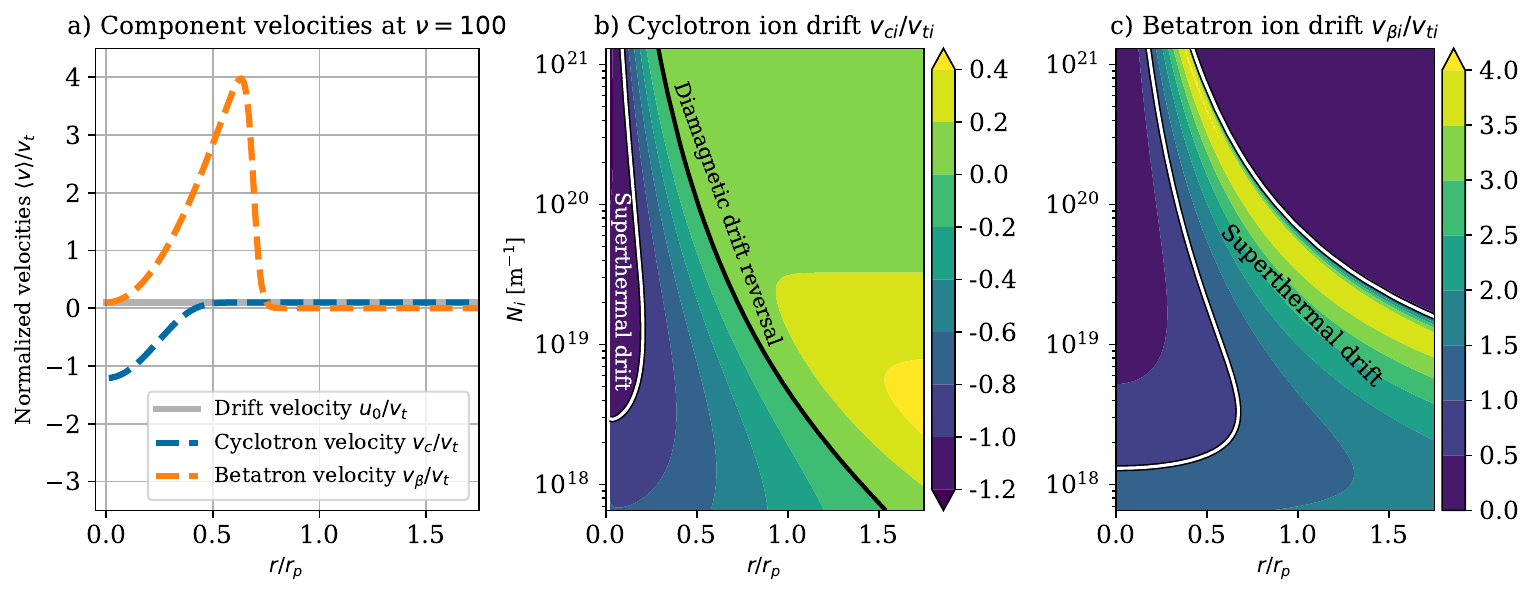}
  \caption{
    Mean velocities of the cyclotron and betatron orbit subpopulations displaying a) component velocities
    at Budker's parameter $\nu=100$ and b-c) the dependence on linear density for ion drifts in an
    electron-deuterium plasma.
    The relative drift between the subpopulations, $v_{\beta i}-v_{c i}$, is superthermal throughout the transitional region.
  \label{fig:component_velocities}}
\end{figure*}

The thickness of the transitional region is defined by where the argument of the function
$\text{erfc}_-(x)$ is sufficiently large, namely $x(r\equiv\Delta)=\sqrt{2}$.
This gives
\begin{equation}\label{eq:region_thickness}
  \frac{\Delta}{\rho_s} = 2\sqrt{\nu_s(e^{(1 + 2\sqrt{\nu_s})/\nu_s}-1)}
  = 2^{3/2}\nu_s^{1/4} + \mathcal{O}(\nu_s^{-1/4}) 
\end{equation}
where $\rho_s$ is the characteristic Larmor radius.
For an electron-deuterium plasma, the ion transitional region thickness is order-of ten Larmor radii for
$\SI{1e19}{m^{-1}}<N_i<\SI{1e24}{m^{-1}}$ (and order-of one hundred beyond).
Nevertheless, the thickness relative to pinch radius accordingly shrinks
as $\Delta / r_p \sim \mathcal{O}(\nu_s^{-1/4})$. 

\subsection{\label{subsec:diamagnetic}Resolving the ``singular current'' of the Z pinch}
A curious element of Z-pinch physics was demonstrated in Ref.~\citenum{haines_1978} that small populations of unmagnetized electrons
and ions conduct what the reference identifies as a ``singular current'' through the magnetic null,
with each species carrying a fraction of the total current ``within one Larmor radius'' of the axis.
The reference showed that ``singular orbits'' on axis carry the plasma current,
while simultaneously the measured current density is distributed throughout the plasma by diamagnetism.

This deep insight appears singular because the derivation is singular; 
it was obtained in the magnetized limit $N\to\infty$ considering the magnetization current
$\vec{\jmath}_M=\nabla\times\big(-\frac{p_\perp}{B^2}\vec{B}\big)$,
\textit{i.e.}{}, the component of diamagnetic current due to the density of magnetic moments.
In the guiding-center limit, current is conducted perpendicular to the magnetic field by the sum of guiding-center drifts
and magnetization current~\cite{northrop_1961}.
Reference~\citenum{haines_1978} reasoned that some flux, the singular current, must cancel the magnetization current at the axis.
This reasoning, while valid asymptotically, requires modification for realistic Z pinches,
especially in the large Larmor radius regime where the density of ion cyclotron orbits is small within the pinch core,
as shown in Figure~\ref{fig:cyclotron_densities}, so that the density of magnetic moments is not equivalent to the plasma density.
Transitional magnetization provides a more complete description at finite $N$ where both cyclotron and betatron populations coexist,
with the singular current of Ref.~\citenum{haines_1978} emerging naturally as a limiting case.

Figures~\ref{fig:cyclotron_densities} and~\ref{fig:cyclotron_currents} 
illustrate that, in regions where the density of cyclotron orbits increases ($dn_c/dr > 0$), the flux $\Gamma_c$
reverses relative to the net species drift flux.
This behavior is simply explained by the equilibrium force balance,
\begin{equation}\label{eq:eql}
  q_sn_s(\vec{E} + \vec{v}_s\times\vec{B}) = \nabla p_s,
\end{equation}
which leads to a general expression for cross-field charge flux,
\begin{equation}\label{eq:general}
  q_sn_s\vec{v}_s = q_sn_s\frac{\vec{E}\times\vec{B}}{B^2} - \frac{\nabla p_s\times\vec{B}}{B^2}.
\end{equation}
This expression is valid for any magnetization regime, but the terms obviously cannot be understood 
in terms of guiding-center dynamics when the constituent orbits do not follow guiding-center dynamics.
A more comprehensive approach emerges from the single-particle motion, suggesting a natural decomposition
$p_s = p_{c,s} + p_{\beta,s}$ into the partial pressures of the cyclotron and betatron populations, from which
\begin{equation}\label{eq:diamagnetic_flux}
  -\frac{\nabla p_s\times\vec{B}}{B^2} = -\frac{\nabla p_{c,s}\times\vec{B}}{B^2} - \frac{\nabla p_{\beta, s}\times\vec{B}}{B^2}.
\end{equation}
The term involving $p_{c,s}$ arises from the guiding-center cyclotron orbit dynamics,
and that with $p_{\beta,s}$ reflects the betatron orbit dynamics.
Thus, within the pinch core where $dp_{c,s}/dr>0$, the diamagnetic particle flux must reverse, which necessitates an opposing
flux of betatron orbits to carry the current.
This confirms the reasoning of Ref.~\citenum{haines_1978} without the singular limit.

Figure~\ref{fig:component_velocities} illustrates how the betatron flux is supported by high-velocity orbits at the layer's edge
by plotting the mean velocities of the two subpopulations (normalized to the thermal speed $v_t$ of the entire population,
whereas the fluxes of Fig.~\ref{fig:cyclotron_currents} are normalized to drift velocity).
Within the transitional region, the relative drift of the subpopulations is superthermal even in the magnetized regime. 
In the magnetized regime, the betatron orbit drift velocity limits to approximately four times the thermal speed at the transitional region edge.

\begin{figure}[bthp]
  \includegraphics[width=\linewidth]{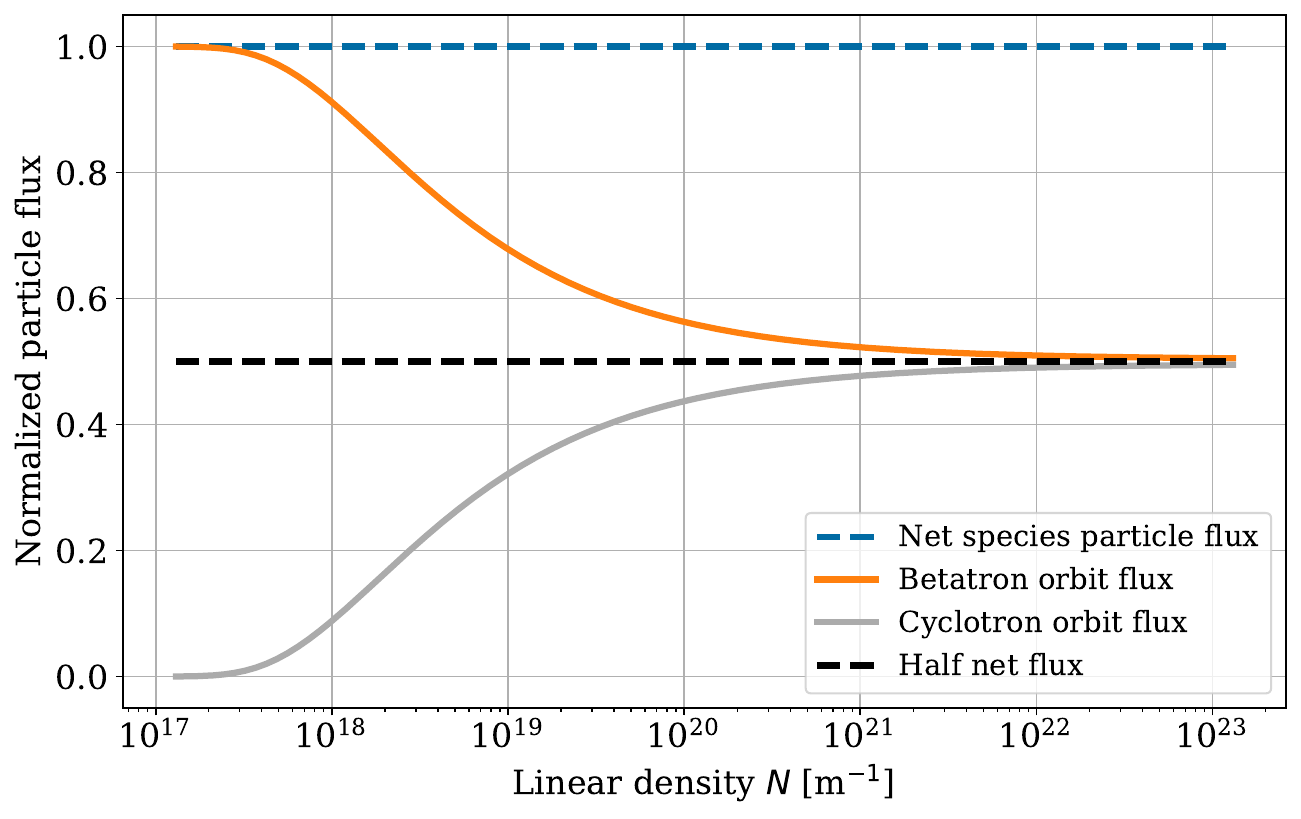}
  \caption{Normalized net ion fluxes in the $\vec{E}^\prime=0$ frame
    ($\int_0^\infty \Gamma_{c,\beta} rdr/\Gamma_0$, integrating the density of Fig.\ref{fig:cyclotron_currents}),
    as a function of linear density $N$ in an electron-deuterium plasma.
    Despite their small number, the betatron orbit flux carries half the ion drift current in the singular limit $N\to\infty$.
    In the large Larmor radius regime, betatron flux carries more than half. 
    \label{fig:net_fluxes}} 
\end{figure}

Figure~\ref{fig:net_fluxes} provides a quantitative demonstration of how the current partition between cyclotron and betatron orbits
varies with linear density.
The formula for the singular current~\cite{haines_1978, haines_2000}, $I_{\text{sing},s} = \frac{4\pi}{\mu_0}\frac{p_s}{j_z}\big|_{r=0}$,
which is valid in the fully magnetized limit, suggests that betatron orbits carry half the total current of the Bennett profile.
But, in the $N\to 0$ limit, unmagnetized orbits must carry the entirety of the current.
Both limits $N\to 0$ and $N\to\infty$ are indeed recovered in Fig.~\ref{fig:net_fluxes}.

In the large Larmor radius regime ($N=10^{18}-10^{20}$ m$^{-1}$), ion betatron orbits
carry a much larger fraction of the current than the asymptotic singular current prediction.
On the other hand, the singular limit safely applies to the electrons in all regimes relevant to the Z pinch.
This allows a stronger statement to be made about the Z-pinch current: at least half of the current, considering both electron and ion drifts,
must be conducted by unmagnetized orbits.
An additional correction to the model of Ref.~\citenum{haines_1978} is that this unmagnetized pinch-core current is not carried within one Larmor radius, but rather
within a much thicker layer given by Eq.~\ref{eq:region_thickness}.

\begin{figure}[h]
\centering

  \includegraphics[width=\linewidth]{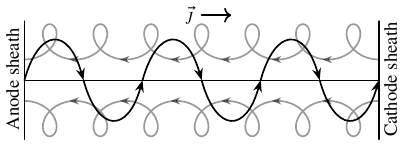}

\caption{
  Illustration of ion fluxes between the electrodes of a static ($\vec{v}_i=0$) pinch plasma, with
  co-current betatron and counter-current cyclotron orbits within the transitional region.
  }
\label{fig:flux_illustration}
\end{figure}

This resolution of the singular current at finite $N$ lends importance to recapitulating the discussion of Ref.~\citenum{haines_1978}
of co- and counter-current ion fluxes in the static Z pinch, sustained by the mean electron motion, $\vec{\jmath} = -ne\vec{v}_e$.
Even though the ion fluid velocity is zero, the subpopulation fluxes are not, as illustrated in Fig.~\ref{fig:flux_illustration}.
In the transitional region, unmagnetized orbits stream co-current, while magnetized orbits stream counter-current.
Thus, in the core, the diamagnetic (magnetized) fluxes of momentum and energy are opposite to
the usual sense in magnetized transport theory~\cite{helander2005collisional}.
The cyclotron and betatron fluxes carry momentum and current through the ends
of the plasma with differing single-particle characteristics~\cite{speiser_1970}.

\subsection{Radial transport considerations in the transitional region\label{subsec:diffusion}}
This section examines classical radial transport while accounting for transitional magnetization.
In magnetized plasmas, transport typically follows from a balance between the single-particle drifts
and the diffusive fluxes due to collisions~\cite{jenko2000, Yoshida_2025}.

In the Z pinch, classical drift-diffusion equilibrium occurs as follows.
The resistive electric field $\vec{E}=\eta\vec{\jmath}$ induces an inward $\vec{E}\times\vec{B}$ drift
in the cyclotron orbit population at velocity $\vec{v}_{E\times B} = \eta(\vec{\jmath}\times\vec{B})/B^2$.
Weighting this drift velocity by the density of cyclotron orbits, $n_c$, 
yields an inward drift flux, $\vec{\Gamma}_{E\times B} = n_c\vec{v}_{E\times B}$.
In the fully magnetized limit, $n=n_c$, this inward flux
is balanced by an outward diffusion flux, $\vec{\Gamma}_\nu=-D_{\perp}\nabla n$,
so that $\vec{\Gamma}_{E\times B}+\vec{\Gamma}_\nu=0$.
A simple calculation recovers classical transport scaling, \textit{i.e.},{}
$D_{\perp}=\eta p/B^2$, or, with the Spitzer formula $\varepsilon_0\eta=\nu/\omega_p^2$, 
the form $D_{\perp}=\nu\rho^2$ where $\rho$ is local Larmor radius and $\nu$ the collision frequency.
Thus, the diffusivity in equilibrium can be estimated with a knowledge of the single-particle drifts.

A key insight of this work is understanding how betatron orbits drift in response to electric fields,
unlike the familiar guiding-center drifts of the cyclotron orbits.
By analyzing the acceleration of a betatron orbit in an axial electric field
(detailed in Appendix~\ref{app:acceleration}), we find its radial $\vec{E}\times\vec{B}$ velocity is related
to its magnetization parameter $\alpha$, Eq.~\ref{eq:alpha_def},
\begin{equation}\label{eq:non_larmor_drift}
  \vec{v}_{\beta,E\times B} = \alpha \vec{v}_{c,E\times B} + \mathcal{O}(\alpha^2).
\end{equation}
This result reveals that betatron orbits have a significantly reduced radial drift response compared to cyclotron orbits.
The ideal betatron orbit with $\alpha \to 0$ has no radial drift response and accelerates freely along the axis,
as in an unmagnetized particle beam.

This response underscores that the betatron orbit, often loosely termed "unmagnetized," is not truly ballistic.
While betatron orbits accelerate in the axial direction in a manner resembling ballistic motion,
their drifts are constrained in the perpendicular plane, somewhat analogous to the cyclotron motion along a field line.
The betatron orbit drift modifies radial transport in the transitional region, where both orbit types coexist,
because the betatron orbits do not undergo the single-particle drifts underlying typical magnetized transport.

Relative to the mean drift, axial acceleration of betatron orbits is asymmetric between co-current and counter-current directions.
Betatron orbits can accelerate indefinitely in the direction of their drift (accumulating $P_z>0$),
whereas counter-current acceleration may transform them into cyclotron orbits (if $P_z<qA_z/2$).
Conversely, cyclotron orbits can become betatron orbits by accumulating sufficient axial momentum (when $P_z>qA_z/2$).
Magnetization transitions occur in this way through dynamic processes like collisions,
with particles exiting a scattering event as magnetized or demagnetized.

\begin{figure}[b]
  \centering
  \resizebox{0.618\columnwidth}{!}{
    \includegraphics[width=\linewidth]{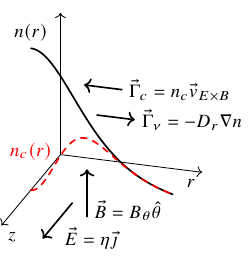}
}
\caption{Classical radial transport in a resistive electric field
  consists of an inward drift flux of guiding centers $\vec{\Gamma}_c$ and a
  counterbalancing diffusion flux $\vec{\Gamma}_\nu$.
  In equilibrium these satisfy $\vec{\Gamma}_c+\vec{\Gamma}_\nu=0$.
  Outside the transitional region, the diffusion coefficient is classical ($D_r\sim |\vec{B}|^{-2}$).}
\label{fig:radial_diffusion}
\end{figure}

The resistive $\vec{E}\times\vec{B}$ drift velocity diverges at the magnetic null, $|v_r|\to \infty$.
This singularity is resolved by a decomposition of the particle density into guiding-center and
non-guiding-center motions, $n = n_c+n_\beta$, as illustrated in Fig.~\ref{fig:radial_diffusion}.
Only the cyclotron orbits drift inwards, but both the cyclotron and betatron orbits diffuse outwards.

Assuming a drift-diffusion equilibrium as in Fig.~\ref{fig:radial_diffusion},
for example at the Pease-Braginskii current~\cite{braginskii_1957, pease_1957}, leads to the estimate
\begin{equation}\label{eq:radial_diffusivity}
  D_{r}(r) = \Big(\frac{n_c}{n}\Big)\nu \rho^2
\end{equation}
as the radial diffusivity, where $\rho=v_t/\omega_c$ is the local
Larmor radius, accounting for the varying magnetic field.
Equation~\ref{eq:radial_diffusivity} limits to a constant value in the core, approaching approximately the 
classical magnetized diffusivity at the edge of the transitional region.
This estimate resolves the singular classical coefficient at the axis,
but assumes drift-diffusion balance maintaining the Bennett equilibrium.
Dynamic effects, non-equilibrium diffusion, and the different thicknesses of the
electron and ion transitional magnetization regions, complicate the diffusive dynamics in the Z-pinch core.

\section{\label{sec:conclusion}Conclusions and implications}
This work has discerned the kinetics of transitional magnetization around the magnetic null through
a detailed analysis of particle orbits in azimuthal self-magnetic fields.
It was shown that general motion in the azimuthal self-field is characterized by two basic frequencies:
the cyclotron frequency $\omega_c$ and the betatron frequency $\omega_\beta$, which motivates the introduction
of a magnetization parameter, $\alpha=\omega_c^2/4\omega_\beta^2$.
Single-particle orbits are classified as either ``cyclotron'' or ``betatron'' orbits based on the dominance of one or the other characteristic frequency.
The magnetization parameter naturally partitions phase space into cyclotron and betatron orbit populations by its relation to the constants of motion.

We find that transitional magnetization typically envelops the axis over tens of Larmor radii and is characterized by the
coexistence of two distinct populations: a guiding-center population of cyclotron orbits, which exhibits a reversed
diamagnetic drift within the envelope, and a counterstreaming population of betatron orbits which supports the axial drift
sustaining the currents.
These drift motions significantly modify classical transport within the transitional magnetization region, a key point of this research.
Our results essentially support the thesis of Ref.~\citenum{haines_1978} on diamagnetism in the Z pinch
and extend that work by treating the betatron orbital effects.
We recover the singular current in the singular limit $N\to\infty$ and resolve this current carried by
unmagnetized orbits into a finite layer around the axis which limits between the singular result and the completely unmagnetized limit.

The cyclotron and betatron orbit subpopulations essentially exhibit distinct symmetry axes.
Cyclotron orbits are symmetric relative to the magnetic field, while the betatron orbits are symmetric
with respect to the electric current.
These symmetry properties, and their implications for the orbit adiabatic invariants, are another key result of this work.
These symmetries enable the development of Chew-Goldberger-Low~\cite{cgl_closure} (CGL)-like double-adiabatic models for the Z pinch,
which is the subject of a future article.

\begin{acknowledgments}
  The authors would like to thank the team at Zap Energy for helpful discussions.
  The information, data, or work presented herein is based in part upon work supported by the
  National Science Foundation under Grant No. PHY-2108419.
\end{acknowledgments}

\section*{Author Declarations}

\subsection*{Conflict of Interest}
The authors have no conflicts to disclose.

\subsection*{Author Contributions}
\noindent\textbf{Daniel W. Crews}: Conceptualization (equal), Investigation (lead),
Writing - original draft (lead), Writing - review and editing (equal).\\
\textbf{Eric T. Meier}: Conceptualization (equal), Supervision (equal),
Writing - review \& editing (equal).\\
\textbf{Uri Shumlak}: Conceptualization (equal), Supervision (equal),
Writing - review \& editing (equal).

\subsection*{Data Availability}
The data that support the findings of this study are available from the corresponding author
upon reasonable request.

\appendix
\section{Effective potential method\label{app:method}}
Potential momentum arises from a charged particle's interaction with a magnetic field,
as expressed by conserved canonical momentum $\vec{P} = m\vec{v} + q\vec{A}$,
where $q\vec{A}$ represents the charge-vector potential coupling~\cite{griffiths_2012}.
Although the magnetic force $q\vec{v}\times\vec{B}$ does no work,
the mechanical momentum evolves as the particle moves.
After reducing dimensions, the exchange of kinetic and potential momentum appears
as an effective potential energy for the remaining kinetic components,
a method commonly used to analyze motion in beams and plasmas
under both external and self-generated magnetic fields~\cite{davidson1998, dodin2006}.

The effective potential arises from the constants of motion for a particle of mass $m$ and charge $q$.
In the frame in which the field is purely magnetic, the three constants are
kinetic energy, axial canonical momentum, and angular momentum, given by
\begin{subequations}\label{eq:constants_of_motion}
\begin{eqnarray}
  H &=& \frac{1}{2}mv_r^2 + \frac{1}{2}mv_\theta^2 + \frac{1}{2}mv_z^2,\label{eq:H}\\
  P_z &=& mv_z + q A_z,\label{eq:Pz}\\
  L_\theta &=& mv_\theta r.\label{eq:Lt}
\end{eqnarray}
\end{subequations}
where the velocity is $\vec{v}=v_r\hat{r} + v_\theta\hat{\theta} + v_z\hat{z}$
in cylindrical coordinates.
Substituting Eqs.~\ref{eq:Pz} and~\ref{eq:Lt} into Eq.~\ref{eq:H} eliminates the velocities $v_\theta$ and $v_z$
to describe one-dimensional motion in an effective potential,
\begin{align}
  H &= K_r + V(r;P_z,L_\theta),\label{eq:Heff}\\
  V(r; P_z, L_\theta) &\equiv \frac{(P_z-qA_z(r))^2}{2m} + \frac{1}{2m}\Big(\frac{L_\theta}{r}\Big)^2,\label{eq:Ueff}
\end{align}
where $K_r \equiv \frac{1}{2}m v_r^2$ is the radial kinetic energy and $V = K_z + K_\theta$
the effective potential energy, which is made up of the axial and azimuthal kinetic energies 
as functions of the constants of motion and of position.

\begin{figure}
  \includegraphics[width=\linewidth]{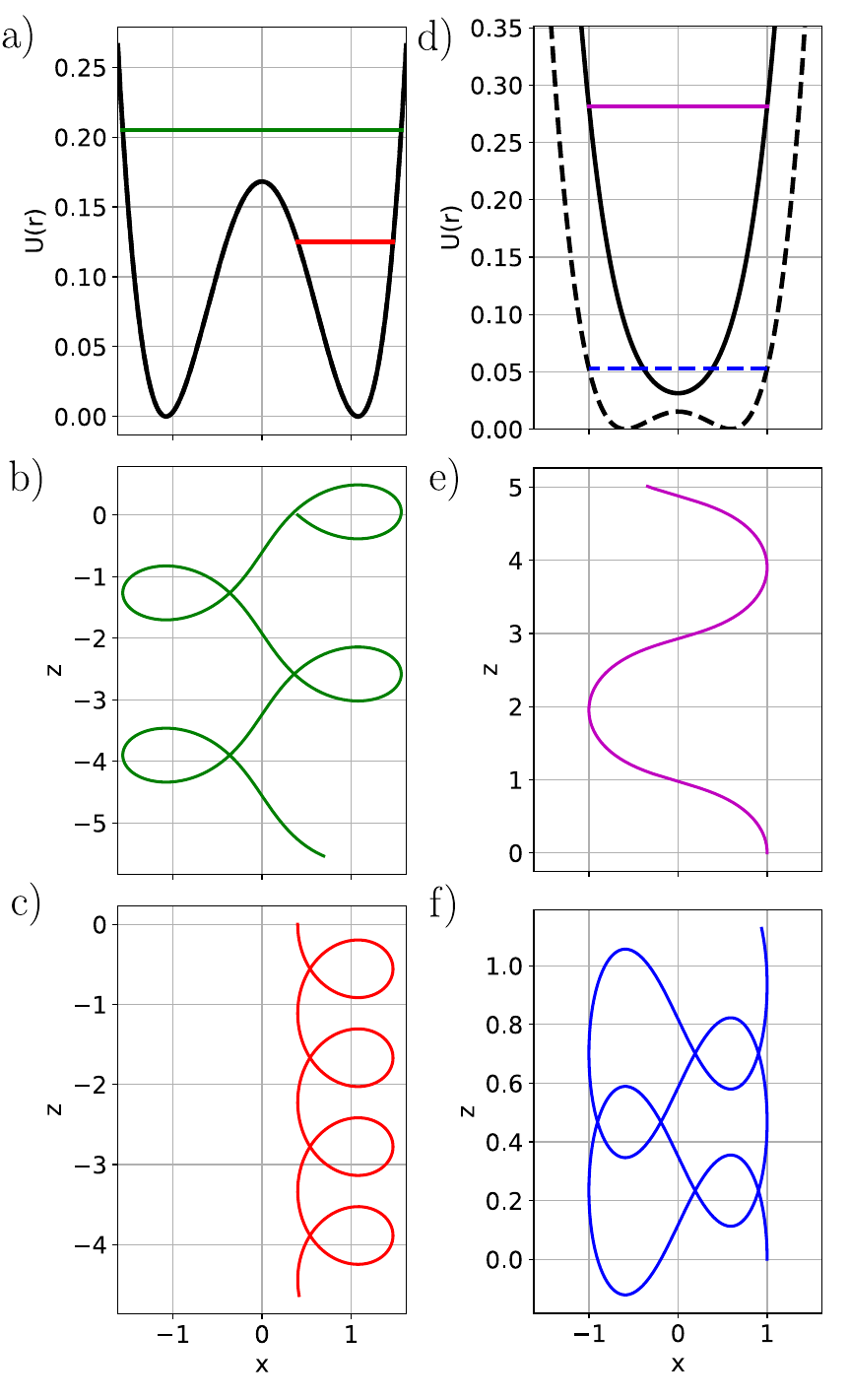}
  \caption{
    Effective potential $V=V(r;P_z)$ (in the $y=0$ plane)
    with meridional trajectories ($L_\theta=0$, $q>0$)
    in the normalized field $B_\theta = x$.
    Low momentum, $P_z<qA_z/2$, and energy, $H<P_z^2/2m$, characterize cyclotron orbits;
    otherwise the orbit is a betatron orbit.
    Orbits (a-c) are backward drifting ($-z$), and (d-f) forward drifting ($+z$).
    Shown are (a-c) cycloidal (cyclotron, in red) and figure-eight (betatron, in green) motions,
    and (d-f) figure-eight (betatron, blue) and snake (betatron, magenta) orbits.
    The momenta shown are (a-c) $P_z=-0.58$, (d, dashed) $P_z=-0.175$, and (d, solid) $P_z=0.25$.
    \label{fig:example_orbits}}
\end{figure}

Figure~\ref{fig:example_orbits} illustrates characteristic orbits in the purely magnetic potential ($L_\theta=0$)
and the simplest azimuthal field $B_\theta = B_0(r/r_p)$.
The sign of canonical momentum controls the effective magnetic potential energy $\frac{(P_z-qA_z(r))^2}{2m}$,
which is of the double-well type for $P_z<0$ (with $q>0$).

\section{Derivation of hybrid guiding-center motion\label{app:guiding_center}}
Finite Larmor radius corrections to the guiding-center motion, incorporating coupling to the betatron oscillation,
are found by expanding the effective potential to second order in Larmor radius $r_1$ relative to distance to the axis $r_0$.
Expand as
\begin{equation}\label{eq:expanded_r}
  r = r_0 + f_1\varepsilon + f_2\varepsilon^2 + \mathcal{O}(\varepsilon^3)
\end{equation}
with $\varepsilon=r_1$ the amplitude (Larmor radius), which acts as the ordering parameter.
The case $r_1\gtrsim r_0$ with $r_1$ small occurs very close to the axis, which is treated in Appendix~\ref{app:analytic}.

The power series of effective potential (Eq.~\ref{eq:Ueff}) for $\delta = r - r_0$ small compared to $r_0$ is
\begin{equation}
  \begin{aligned}
    \frac{V}{m} = &\frac{K_{z0} + K_{\theta 0}}{m} + (v_{z0}\omega_c-r_0\omega_\beta^2)\delta\\
                              &+\frac{1}{2}(\omega_c^2 + v_{z0}\omega_c' + 3\omega_\beta^2)\delta^2\\
    &+\frac{1}{6}(v_{z0}\omega_c'' + 3\omega_c'\omega_c - 12r_0^{-1}\omega_\beta^2)\delta^3 + \mathcal{O}(\delta^4),
  \end{aligned}
\end{equation}
where prime denotes a derivative evaluated at $r=r_0$.
Collecting by order in $\varepsilon$ in $\ddot{r}=-m^{-1}\frac{dV}{dr}|_{r=r_0}$ gives,
\begin{alignat}{2}
  \mathcal{O}(\varepsilon^0):\quad & \ddot{r}_0&=& -(v_{z0}\omega_c - r_0\omega_\beta^2),\\
  \mathcal{O}(\varepsilon^1):\quad &\ddot{f}_1&=& -\omega^2f_1,\\
  \mathcal{O}(\varepsilon^2):\quad &\ddot{f}_2&=& -\omega^2f_2 - \mathcal{A}f_1^2,
\end{alignat}
where $\mathcal{A} \equiv \frac{1}{2}(v_{z0}\omega_c'' - 3\omega_c\omega_c' - 12r_0^{-1}\omega_\beta^2)$
and the frequency $\omega^2 \equiv \omega_c^2 + v_{z0}\omega_c' + 3\omega_\beta^2$.
The zeroth-order balance with $\ddot{r}_0=0$ describes the axis-encircling curvature drift orbit,
on top of which the first-order motion is as described in Sec.~\ref{subsubsec:intro_linearized},
with $f_1 = \sin(\omega t)$.

To second-order in $\varepsilon$, the solution satisfying initial conditions $f_2(0)=f_2'(0)=0$ is
\begin{equation}
  f_2 = -\frac{\mathcal{A}}{2\omega^2}\Big(1 - \frac{4}{3}\cos(\omega t) +
  \frac{2}{3}\sin(\omega t) + \frac{1}{6}\cos(2\omega t)\Big).
\end{equation}
Gyro-averaging Eq.~\ref{eq:expanded_r} (over $T = 2\pi/\omega$) gives
\begin{equation}\label{eq:avg_radius}
  \langle r\rangle = r_0 - \frac{\mathcal{A}}{2\omega^2}r_1^2 + \mathcal{O}(r_1^3)
\end{equation}
and can be expressed as Eq.~\ref{eq:displacement}.
Corrections to the guiding-center drift follow from expanding the vector potential to second order,
\begin{equation}
  A_z = A_{z0} - B_{\theta 0}\delta - \frac{1}{2} B_{\theta 0}'\delta^2 + \mathcal{O}(\delta^3),
\end{equation}
and combining with $P_z = mv_z + qA_z$ to obtain
\begin{equation}\label{eq:axial_velocity}
  v_z = v_{z0} + \omega_c\delta + \frac{1}{2}\omega_c'\delta^2 + \mathcal{O}(\delta^3).
\end{equation}
Averaging Eq.~\ref{eq:axial_velocity} over the gyroperiod produces Eq.~\ref{eq:drift1}.

\section{Exact solutions in radially uniform current\label{app:analytic}}
Around the magnetic null, current density is approximately uniform and the azimuthal magnetic field increases
linearly with radius.
In this situation, exact solutions for orbits may be developed including angular momentum
because the vector potential is simply quadratic, \textit{i.e.},{}
$A_z = -\frac{1}{2}A_0(r/\ell)^2$ with $A_0=B_0\ell$, $\ell$ a scale length, and $B_0$ characteristic flux density.

A standard form is obtained from Eq.~\ref{eq:Heff} as 
\begin{equation}\label{eq:mult_rsq}
  \frac{1}{4}\Big(\frac{dr^2}{dt}\Big)^2 = r^2(v^2 - (p_z - a_z)^2) - \ell_\theta^2
\end{equation}
where $v^2\equiv 2H/m$, $a_z \equiv qA_z/m$, $p_z \equiv P_z/m$, and $\ell_\theta\equiv L_\theta / m$.
With $a_z\sim r^2$, the solutions to Eq.~\ref{eq:mult_rsq} can be expressed with elliptic functions.

Let time be normalized to the characteristic gyroperiod $\omega_{c0}^{-1}=(qB_0/m)^{-1}$,
length to $\ell$, both speed $v$ and vector potential $a_z$ to $\ell\omega_{c0}$,
and define normalized time as $\tau \equiv \omega_{c0}t/\sqrt{2}$.
Equation~\ref{eq:mult_rsq} rearranges to
\begin{equation}\label{eq:diff_eq_for_A}
  \Big(\frac{da_z}{d\tau}\Big)^2 = 4\Big(a_z^3 - 2p_za_z^2 + (p_z^2 - v^2)a_z - \frac{\ell_\theta^2}{2}\Big)
\end{equation}
where the variables are all normalized as described.
Equation~\ref{eq:diff_eq_for_A} is transformed to Weierstrass normal
form by changing variables to a depressed cubic
with $Q \equiv a_z - 2p_z/3$ and $N = p_z/3$, which gives
\begin{equation}\label{eq:depressed_cubic}
\Big(\frac{dQ}{d\tau}\Big)^2 = 4Q^3 - g_2Q - g_3
\end{equation}
where the parameters $g_2$, $g_3$ are defined as 
\begin{subequations}
\begin{eqnarray}
g_2&=&4(v^2 + 3N^2), \label{eq:g2}
\\
g_3&=&8N(v^2-N^2) + 2\ell_\theta^2. \label{eq:g3}
\end{eqnarray}
\end{subequations}
The general solution of Eq.~\ref{eq:depressed_cubic} is
the Weierstrass $\wp$-function~\cite{pastras2020weierstrass}
\begin{equation}\label{eq:wp_sol}
  Q = \wp(\tau + \tau_0;g_2,g_3)
\end{equation}
with $\tau_0$ chosen such that $\wp(\tau_0;g_2,g_3)=Q_0$ at $t=0$.
The properties of the solution are controlled by the roots $(e_1,e_2,e_3)$ with $e_1>e_2>e_3$
of the polynomial $\mathcal{P}(Q) = 4Q^3 - g_2Q - g_3$.

Unordered as $(r_1, r_2, r_3)$, the roots take the form
\begin{subequations}
  \begin{eqnarray}
    r_{1,2} &=& \frac{p_z}{3} - \dot{z}_0(\alpha \mp \Delta), \label{eq:e1}
    \\
    r_3 &=& \frac{p_z}{3} - \dot{z}_0, \label{eq:e3}
    \\
    \Delta &\equiv& \sqrt{(1+\alpha)^2 + K_{\theta 0}/K_{z0}} \label{eq:delta}
\end{eqnarray}
\end{subequations}
where $\dot{z}_0$ is the initial $z$-velocity,
$\Delta$ is a quadratic discriminant, and $\alpha=\omega_c^2/4\omega_\beta^2$
is the magnetization parameter.  
The discriminant measures the relative importance of the magnetization parameter
and the initial parallel-to-perpendicular kinetic energy ratio where
$K_{\theta 0} \equiv (L_\theta/r_0)^2/2m$ and $K_{z0} \equiv m\dot{z}_0^2/2$.

Equation~\ref{eq:wp_sol} is expressed with Jacobi elliptic functions as
\begin{equation}\label{eq:radial_solution_L}
  \frac{1}{2}r^2(t) = \frac{1}{2}r_0^2 - C\text{sn}^2(\omega t | m)
\end{equation}
where the elliptic modulus $m \equiv (e_2 - e_3) / (e_1 - e_3)$, the amplitude $C \equiv e_2 - e_3$,
and the frequency $\omega \equiv \sqrt{(e_1-e_3)/2}$.
In use, the roots $(r_1,r_2,r_3)$ must first be ordered into $e_1>e_2>e_3$,
which generally depends on the sign of $p_z$, etc.

\begin{figure}[tpbh]
  \includegraphics[width=\linewidth]{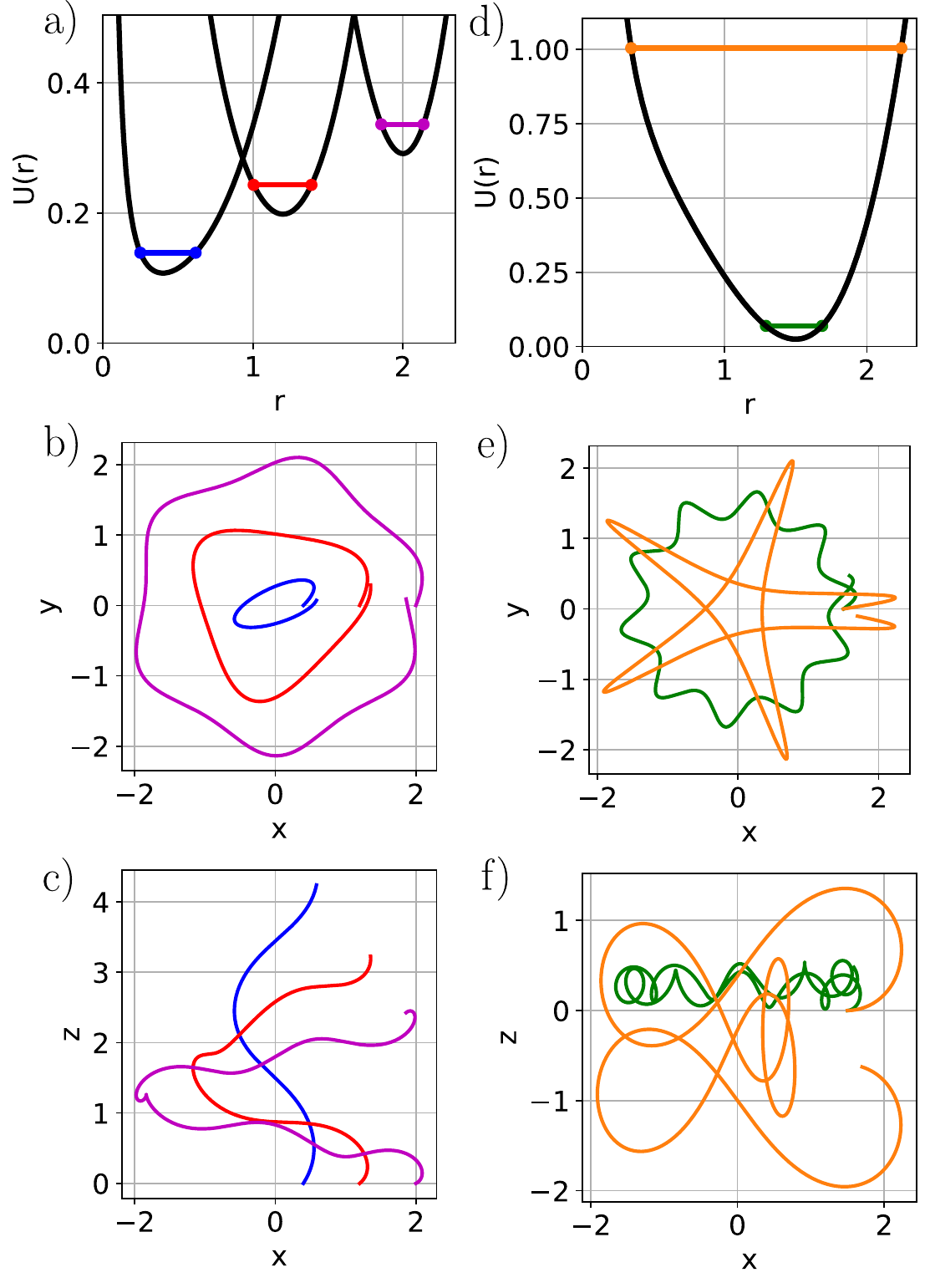}
  \caption{
    A variety of axis-encircling orbits, with a)-c) illustrating the progression from betatron
    to cyclotron orbit in blue, red, and magenta.
    Parts d)-f) contrast a low-energy cyclotron orbit (green) to a high-energy betatron orbit (orange).
    Parts a)-c) have magnetization parameters of
    $\alpha=0.1$, $1.5$, and $7.1$ for the blue, red, and magenta orbits respectively.
    \label{fig:helical_orbits}}
\end{figure}

When $K_{\theta 0}=0$, the roots reduce to $e_{1,3} = N \pm \dot{z}_0$ and
$e_2 = N - \dot{z}_0(1+2\alpha)$ which describe meridional orbits in the $(r,z)$ plane
(as in Fig.~\ref{fig:example_orbits}).
These are the solutions in the planar current sheet historically presented by
Speiser~\cite{speiser_1965a} and Sonnerup~\cite{sonnerup_1971}, and are not repeated here.
The elliptic modulus of the meridional orbits is the magnetization parameter, $m=\alpha$.
Figure~\ref{fig:helical_orbits} illustrates the kinds of orbits described by the solutions
of this section, visualizing how transitional magnetization is governed by the relative rates
of azimuthal circulation and radial oscillation.

\subsection{Standard guiding-center motion at large magnetization}
Standard guiding-center drifts are recovered to leading-order in
large magnetization parameter $\alpha \gg 1$
upon gyroaveraging the motion,
for which a useful identity is
\begin{equation}\label{eq:averaged_sine_squared}
  \frac{1}{K(m)}\int_0^{K(m)}\text{sn}^2(t|m)dt =
  \frac{1}{m}\Big(1 - \frac{E(m)}{K(m)}\Big).
\end{equation}
The axial drift is computed using Eqs.~\ref{eq:radial_solution_L} and~\ref{eq:averaged_sine_squared} to yield
\begin{equation}\label{eq:avg_z_velocity}
  \langle\dot{z}\rangle = \dot{z}_0\Big(1 - (1 - \alpha + \Delta)\Big(1 - \frac{E(m)}{K(m)}\Big)\Big).
\end{equation}
Expanding terms as
\begin{subequations}
  \begin{eqnarray}
    m &=& 1 - 2\alpha^{-1} + \mathcal{O}(\alpha^{-2}), \label{eq:exp1}
    \\
    1 - \frac{E(m)}{K(m)} &=& \frac{1 - 2\alpha^{-1}}{2} + \mathcal{O}(\alpha^{-2}),\label{eq:exp2}
    \\
    1 - \alpha + \Delta &=& 2 + \frac{K_{\parallel 0}}{2K_{\perp 0}}\alpha^{-1} + \mathcal{O}(\alpha^{-2}),\label{eq:exp3}
\end{eqnarray}
\end{subequations}
and substituting Eqs.~\ref{eq:exp1}-\ref{eq:exp3} into Eq.~\ref{eq:avg_z_velocity} gives
\begin{equation}\label{eq:guiding-center-spin-appendix}
  \langle\dot{z}\rangle = \frac{2 K_{\theta 0}}{r_0B_{\theta 0}} - \frac{K_{z0}|\nabla B_{\theta 0}|}{B_{\theta 0}^2} + \mathcal{O}(\alpha^{-2}).
\end{equation}
The first term is the curvature drift and the second the $\nabla|\vec{B}|$-drift,
which arise for these large orbits without further corrections because the field gradient
is taken to be constant for all radii.
This demonstrates that $\alpha$ is the key control parameter of magnetization
even for large Larmor radius orbits.

\section{Acceleration and drift of betatron orbits\label{app:acceleration}}
In the magnetized limit ($\alpha\gg 1$), guiding centers drift at the 
$\vec{E}\times\vec{B}$ velocity in response to an electric field,
\begin{equation}\label{eq:exb}
  \vec{v}_{E\times B} = \frac{\vec{E}\times\vec{B}}{B^2}.
\end{equation}
Equation~\ref{eq:exb} breaks down, however, for the resistive field $\vec{E}=\eta\vec{\jmath}$
at the magnetic null which predicts $|\vec{v}_{E\times B}|\to \infty$.

\begin{figure}[t]
  \centering
  
  
  

  
  \includegraphics[width=0.555\linewidth]{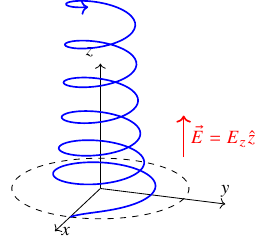}
  \caption{Acceleration of a betatron orbit in an axial electric field, in the
    limit of small magnetization parameter ($\alpha\ll 1$).
    Acceleration is ballistic in the axial direction,
    increasing the betatron frequency as $\omega_\beta^2 = (\omega_c/r)v_z$.
    As the betatron frequency increases, the orbital radius pulls inward through conservation of
    angular momentum, satisfying $r^4v_z=\text{const}$.} 
  \label{fig:helix2}
\end{figure}

To resolve this singularity, it is necessary to understand the response of the betatron orbits to
an axial electric field.
In the ideal betatron limit ($\alpha\ll 1$), the terms involving Larmor gyration in the equation of motion
(Eq.~\ref{eq:linear_motion}) may be neglected.
The trajectory simply accelerates freely in the axial direction.
Consequently, the betatron frequency increases as the square-root of the velocity,
$\omega_\beta^2\sim v_z$, and to conserve angular momentum the orbit drifts radially inward
satisfying the invariant $r^4v_z=\text{const}$.{},
from which the inward drift is calculated as
\begin{equation}
  \vec{v}_{\beta,\text{drift}} = \Big(\frac{q}{m}\Big)^2\frac{\vec{E}\times\vec{B}}{4\omega_\beta^2}
\end{equation}
which leads to Eq.~\ref{eq:non_larmor_drift} by introducing the parameter
$\alpha = \omega_c^2/4\omega_\beta^2$.
Figure~\ref{fig:helix2} depicts this acceleration process.

\bibliography{orbite_rev}

\end{document}